\documentclass[letterpaper,twocolumn,10pt]{article}
\usepackage{usenix2019_v3}

\usepackage{algorithm}
\usepackage{algpseudocode}
\usepackage{array}
\usepackage{amsmath}
\usepackage{amssymb}
\usepackage{amsthm}
\usepackage{booktabs}
\usepackage{calc}
\usepackage{cleveref}
\usepackage{colortbl}
\usepackage{mathtools}
\usepackage{multirow}
\usepackage{tikz}
\usetikzlibrary{calc, positioning}
\usepackage{pgfplots}
\pgfplotsset{compat=1.18}
\usepackage{subcaption}
\usepackage{makecell}
\usepackage{url}
\usepackage{footnote}
\usepackage{dblfloatfix}
 
\definecolor{supergreen}{RGB}{0,150,0}

\algrenewcommand\algorithmicrequire{\textbf{Input:}}
\algrenewcommand\algorithmicensure{\textbf{Output:}}

\algblock{ParFor}{EndParFor}
\algnewcommand\algorithmicparfor{\textbf{parallel for}}
\algnewcommand\algorithmicpardo{\textbf{do}}
\algnewcommand\algorithmicendparfor{\textbf{end\ for}}
\algrenewtext{ParFor}[1]{\algorithmicparfor\ #1\ \algorithmicpardo}
\algrenewtext{EndParFor}{\algorithmicendparfor}

\newcolumntype{P}[1]{>{\centering\arraybackslash}p{#1}}

\newcommand{\innerproduct}[2]{\langle #1, #2 \rangle}
\DeclarePairedDelimiter{\ceil}{\lceil}{\rceil}
\DeclarePairedDelimiter{\floor}{\lfloor}{\rfloor}

\newlength{\DepthReference}
\newlength{\HeightReference}
\newlength{\Width}%
\newcommand{\MyColorBox}[2][red]%
{%
    \settowidth{\Width}{#2}%
    \colorbox{#1}%
    {%
        \raisebox{-\DepthReference}%
        {%
                \parbox[b][\HeightReference+\DepthReference][c]{\Width}{\centering#2}%
        }%
    }%
}

\makeatletter
\newcommand\footnoteref[1]{\protected@xdef\@thefnmark{\ref{#1}}\@footnotemark}
\makeatother

\newcommand{\mytilde}{\raisebox{0.5ex}{\texttildelow}}

\DeclareMathOperator{\C}{\mathbb{C}}
\DeclareMathOperator{\N}{\mathbb{N}}
\DeclareMathOperator{\R}{\mathbb{R}}
\DeclareMathOperator{\Z}{\mathbb{Z}}

\DeclareMathOperator{\cmp}{Cmp}
\DeclareMathOperator{\ind}{Ind}
\DeclareMathOperator{\eq}{Eq}

\newcommand{\maskR}{\textsf{MaskR}}
\newcommand{\maskC}{\textsf{MaskC}}
\newcommand{\sumR}{\textsf{SumR}}
\newcommand{\sumC}{\textsf{SumC}}
\newcommand{\replR}{\textsf{ReplR}}
\newcommand{\replC}{\textsf{ReplC}}
\newcommand{\transR}{\textsf{TransR}}
\newcommand{\transC}{\textsf{TransC}}

\newcommand{\rank}{\textsf{Rank}}

\newcommand{\vectorLength}{N}
\renewcommand{\vector}{v}
\newcommand{\encVector}{V}
\newcommand{\comparison}{c}
\newcommand{\encComparison}{C}
\newcommand{\ranking}{r}
\newcommand{\encRanking}{R}
\newcommand{\orderStat}{o}
\newcommand{\encOrderStat}{O}
\newcommand{\sortedVec}{s}
\newcommand{\encSortedVec}{S}
\newcommand{\sortingMask}{m}
\newcommand{\encSortingMask}{M}
\newcommand{\encBlock}{V}
\newcommand{\numBlocks}{L}
\newcommand{\blockSize}{B}
\newcommand{\approximationDegree}{d}
\newcommand{\approximationDepth}{D}
\newcommand{\offsetVec}{f}
\newcommand{\encOffsetVec}{F}
\newcommand{\equality}{e}
\newcommand{\encEquality}{E}
\newcommand{\scalingEquality}{u}
\newcommand{\encScalingEquality}{U}
\newcommand{\tieSize}{t}
\newcommand{\encTieSize}{T}

\newcommand{\partitionLower}{\mathcal{L}}
\newcommand{\partitionEqual}{\mathcal{E}}
\newcommand{\partitionGreater}{\mathcal{G}}
\newcommand{\partitionEqualScaled}{\mathcal{U}}

\newcommand{\tabcolsepValue}{1.2pt}
\newcommand{\arraystretchValue}{1.38}

\begin{document}

\date{}

\title{\Large \bf Efficient Ranking, Order Statistics, and Sorting under CKKS}


\author{
{\rm Federico Mazzone}\\
University of Twente
\and
{\rm Maarten Everts}\\
University of Twente\\Linksight
\and
{\rm Florian Hahn}\\
University of Twente
\and
{\rm Andreas Peter}\\
Carl von Ossietzky\\Universität Oldenburg
} 

\maketitle

\begin{abstract}
Fully Homomorphic Encryption (FHE) enables operations on encrypted data, making it extremely useful for privacy-preserving applications, especially in cloud computing environments.
In such contexts, operations like ranking, order statistics, and sorting are fundamental functionalities often required for database queries or as building blocks of larger protocols.
However, the high computational overhead and limited native operations of FHE pose significant challenges for an efficient implementation of these tasks.
These challenges are exacerbated by the fact that all these functionalities are based on comparing elements, which is a severely expensive operation under encryption.

Previous solutions have typically based their designs on swap-based techniques, where two elements are conditionally swapped based on the results of their comparison.
These methods aim to reduce the primary computational bottleneck: the \textit{comparison depth}, which is the number of non-parallelizable homomorphic comparisons in the algorithm.
The current state of the art solutions for sorting by Lu et al.~(IEEE S\&P'21) and Hong et al.~(IEEE TIFS 2021), for instance, achieve a comparison depth of $\log^2{\vectorLength}$ and $k \log_k^2{\vectorLength}$, respectively.

In this paper, we address the challenge of reducing the comparison depth by shifting away from the swap-based paradigm.
We present solutions for ranking, order statistics, and sorting, that achieve a comparison depth of up to 2 (constant), making our approach highly parallelizable and suitable for hardware acceleration.
Leveraging the SIMD capabilities of the CKKS FHE scheme, our approach re-encodes the input vector under encryption to allow for simultaneous comparisons of all elements with each other.
The homomorphic re-encoding incurs a minimal computational overhead of $O(\log{\vectorLength})$ rotations.
Experimental results show that our approach ranks a 128-element vector in approximately 5.76s, computes its argmin/argmax in 12.83s, and sorts it in 78.64s.
\end{abstract}


\section{Introduction}
\label{sec:introduction}

Fully Homomorphic Encryption (FHE) is a cryptographic primitive that enables performing unbounded operations on encrypted data, without decrypting them first.
It is a fundamental building block for designing non-interactive protocols in privacy-preserving applications, and can be used to maintain the confidentiality of data stored in the cloud, while enabling outsourced computations on it.
Despite the effort of recent developments to make this technology more efficient and closer to be usable in real-world applications \cite{smart2010fully,brakerski2014efficient,brakerski2014leveled,brakerski2011fully}, computing under FHE is still problematic due to both its serious computational overhead and limited native operations.
In particular, it is challenging to realize even fundamental functions efficiently, like \textit{ranking}, computing \textit{order statistics}, and \textit{sorting}, which are frequently required database operations.
These functionalities also find applications in diverse fields, for instance in privacy-preserving machine learning, where they can be used to evaluate max-pooling layers and the argmax output layer in neural networks~\cite{jovanovic2022private,zhang2024secure}, or to perform private inference of decision trees~\cite{lu2018non,tueno2020non}.

\begin{savenotes}
\begin{table*}[]
    \centering
    \caption{Summary of related work.}
    \label{tab:related_work}
    \begin{tabular}{lllllp{116pt}}
        \toprule
        \thead{Paper} & \thead{Function} & \thead{Comp. \\ Depth} & \thead{Comput. \\ Complexity} & \thead{FHE \\ Scheme} & \thead{Remarks} \\
        \midrule
        \makecell[tl]{Chatterjee et al.~\cite{chatterjee2013accelerating} \\ (Indocrypt 2013)} & \makecell[tl]{Bubble Sort, \\ Insertion Sort} & \makecell[tl]{$\vectorLength^2$ \\ $\vectorLength^2$} & \makecell[tl]{$O(\vectorLength^2)$ \\ $O(\vectorLength^2)$} & \makecell[tl]{SV \\ SV} & Tested up to 40 elements, for which it runs in around 359.42 minutes (Bubble Sort) and 362.62 minutes (Insertion Sort). \\
        \makecell[tl]{Chatterjee et al.~\cite{chatterjee2017sorting} \\ (IEEE TSC 2017)} & Quick Sort & $\vectorLength^2$ & $O(\vectorLength^2)$ & SV & Tested up to 40 elements, for which it runs in around 779.28 minutes. \\
        \makecell[tl]{Emmadi et al.~\cite{emmadi2015updates} \\ (ICCCRI 2015)} & \makecell[tl]{Bitonic Sort, \\ Odd-Even Merge Sort} & \makecell[tl]{$\log^2{\vectorLength}$ \\ $\log^2{\vectorLength}$} & \makecell[tl]{$O(\vectorLength\log^2{\vectorLength})$ \\ $O(\vectorLength\log^2{\vectorLength})$} & \makecell[tl]{SV \\ SV} & Tested up to 64 elements, for which it runs in around 52.63 minutes (Bitonic Sort) and 42.65 minutes (Odd-Even Merge Sort). \\
        \makecell[tl]{Lu et al.~\cite{lu2021pegasus} \\ (IEEE S\&P 2021)} & \makecell[tl]{Bitonic Sort \\ (switching to FHEW)} & $\log^2{\vectorLength}$ & $O(\vectorLength \log^2{\vectorLength})$ & CKKS & Tested up to 64 elements, for which it runs in 409.09s in a 4-thread setting (without taking into account the scheme switching cost). \\
        \makecell[tl]{Hong et al.~\cite{hong2021efficient} \\ (IEEE TIFS 2021)} & k-Way Sorting Networks & $k \log_k^2{\vectorLength}$ & $O(\vectorLength k \log_k^2{\vectorLength})$ & CKKS & It takes around 87.35 minutes to sort 625 elements. \\
        \makecell[tl]{Jovanovic et al.~\cite{jovanovic2022private} \\ (ACM CCS 2022)} & Argmax & $\vectorLength$ & $O(\vectorLength)$ & CKKS & It computes the argmax of 128 elements in approximately 92.31 minutes. \\
        \makecell[tl]{Zhang et al.~\cite{zhang2024secure} \\ (NDSS 2025)} & Argmax & $\log{\vectorLength} + 1$ & $O(\log{\vectorLength})$ & CKKS & It computes the argmax of 128 elements in approximately 5.05 minutes. \\
        \makecell[tl]{Our work} & \makecell[tl]{Ranking \\ k-Statistics (incl. Argmax) \\ Sorting} & \makecell[tl]{$\textbf{1}$\footnotemark \\ $\textbf{2}$\footnoteref{fn:mult-depth} \\ $\textbf{2}$\footnoteref{fn:mult-depth}} & \makecell[tl]{$O(\numBlocks^2)$ \\ $O(\numBlocks^2)$ \\ $O(\numBlocks^2)$ \\ with $\numBlocks = \vectorLength / \blockSize$ \\ $\blockSize \in \{128, 256\}$} & \makecell[tl]{CKKS \\ CKKS \\ CKKS} & Tested up to 16384 elements. It ranks 128 elements in 5.76s, computes their argmin/ argmax in 12.83s, and sorts them in 78.64s.
        \\
        \bottomrule
    \end{tabular}
\end{table*}
\end{savenotes}

\footnotetext{\label{fn:mult-depth}
Precisely, in terms of multiplicative depth, ranking requires up to $\approximationDepth_C + 4$ levels, while the extraction of $k$-statistics and sorting require up to $\approximationDepth_C + \approximationDepth_I + 4$ and $\approximationDepth_C + \approximationDepth_I + 6$ levels, respectively. Here, $\approximationDepth_C$ and $\approximationDepth_I$ represent the multiplicative depth of the comparison and indicator circuits (see \Cref{sec:main-design}).
}

Several works in the literature have attempted to implement these functionalities efficiently (see \Cref{tab:related_work}).
Many studies have focused on sorting under both the Smart-Vercauteren (SV) scheme~\cite{smart2010fully} and the 
Cheon-Kim-Kim-Song scheme (CKKS), which is particularly relevant as it enables floating-point arithmetic on vectors of data in a Single Instruction Multiple Data (SIMD) fashion.
These approaches typically implement swap-based sorting methods, where at each round two elements are compared and conditionally swapped~\cite{chatterjee2013accelerating,chatterjee2017sorting,emmadi2015updates,lu2021pegasus,hong2021efficient}.
Similarly, other works have focused on computing the argmax of a vector of elements encrypted in a CKKS ciphertext, also relying on swap-based techniques~\cite{jovanovic2022private,zhang2024secure}.
However, comparing two values under encryption is significantly expensive, resulting in the bottleneck of these methods being the \textit{comparison depth}.
The comparison depth refers to the number of comparisons that must be executed in series, and hence cannot be parallelized.
Consequently, \textbf{the main problem lies in designing algorithms capable of achieving a low comparison depth}.
This task is made particularly challenging by the fact that any swap-based algorithm will have worst case complexity under encryption~\cite{emmadi2015updates}.

In this paper, we design and implement novel algorithms for the aforementioned functionalities which, for the first time, require a constant comparison depth of 2 only.
To overcome the limitations of previous solutions, we avoid relying on the strategy of swapping elements under encryption.
Our approach heavily exploits the SIMD capabilities of CKKS.
The core idea is to use suitable homomorphic rotations and masking to re-encode the encrypted elements in such a way that allows us to compare all elements against each other simultaneously.
By aggregating the outcome of this comparison, we compute the rank of each element within the vector.
Then, by leveraging appropriate indicator functions, it is possible to use the computed ranks to extract any order statistic and their position, including minimum, maximum, and median (or any percentile).
Finally, we show how to parallelize this extraction process to implement a sorting functionality.

By employing only recursive rotation-based operations, we make sure that the re-encoding under encryption requires only $O(\log{\vectorLength})$ rotations, where $\vectorLength$ is the vector length, thus causing minimal computational overhead.
Moreover, the low comparison depth makes our solution highly parallelizable, thus suitable for potential hardware acceleration, such as on GPUs.
While we leave this direction to future research, in our present work we show how to further optimize our solution algorithmically in multithreaded environments.
Importantly, our approach is agnostic to the specific implementation of the comparison function.
Even with a basic implementation, our approach can rank a vector of 128 elements in approximately 5.76s, compute its argmin/argmax in 12.83s, and sort it in 78.64s.



\section{Preliminaries}

\begin{figure*}[!htb]
    \centering
    \resizebox{\textwidth}{!}{%
    \begin{tabular}{ccccc}
    \renewcommand{\arraystretch}{1.2}
    \renewcommand{\tabcolsep}{2pt}
    \begin{tabular}{|P{9pt}|P{9pt}|P{9pt}|P{9pt}|P{9pt}|P{9pt}|P{9pt}|P{9pt}|}
        \hline
        \cellcolor{blue!10}$\vector_1$ & 0 & 0 & 0 & 0 & 0 & 0 & 0 \\ \hline
        \cellcolor{blue!10}$\vector_2$ & 0 & 0 & 0 & 0 & 0 & 0 & 0 \\ \hline
        \cellcolor{blue!10}$\vector_3$ & 0 & 0 & 0 & 0 & 0 & 0 & 0 \\ \hline
        \cellcolor{blue!10}$\vector_4$ & 0 & 0 & 0 & 0 & 0 & 0 & 0 \\ \hline
        \cellcolor{blue!10}$\vector_5$ & 0 & 0 & 0 & 0 & 0 & 0 & 0 \\ \hline
        \cellcolor{blue!10}$\vector_6$ & 0 & 0 & 0 & 0 & 0 & 0 & 0 \\ \hline
        \cellcolor{blue!10}$\vector_7$ & 0 & 0 & 0 & 0 & 0 & 0 & 0 \\ \hline
        \cellcolor{blue!10}$\vector_8$ & 0 & 0 & 0 & 0 & 0 & 0 & 0 \\ \hline
    \end{tabular}
    &
    \renewcommand{\arraystretch}{1.2}
    \renewcommand{\tabcolsep}{2pt}
    \begin{tabular}{|P{9pt}|P{9pt}|P{9pt}|P{9pt}|P{9pt}|P{9pt}|P{9pt}|P{9pt}|}
        \hline
        $\vector_1$ & 0 & 0 & 0 & \cellcolor{blue!10}$\vector_5$ & 0 & 0 & 0 \\ \hline
        $\vector_2$ & 0 & 0 & 0 & \cellcolor{blue!10}$\vector_6$ & 0 & 0 & 0 \\ \hline
        $\vector_3$ & 0 & 0 & 0 & \cellcolor{blue!10}$\vector_7$ & 0 & 0 & 0 \\ \hline
        $\vector_4$ & 0 & 0 & 0 & \cellcolor{blue!10}$\vector_8$ & 0 & 0 & 0 \\ \hline
        $\vector_5$ & 0 & 0 & 0 & \cellcolor{blue!10}$\vector_1$ & 0 & 0 & 0 \\ \hline
        $\vector_6$ & 0 & 0 & 0 & \cellcolor{blue!10}$\vector_2$ & 0 & 0 & 0 \\ \hline
        $\vector_7$ & 0 & 0 & 0 & \cellcolor{blue!10}$\vector_3$ & 0 & 0 & 0 \\ \hline
        $\vector_8$ & 0 & 0 & 0 & \cellcolor{blue!10}$\vector_4$ & 0 & 0 & 0 \\ \hline
    \end{tabular}
    &
    \renewcommand{\arraystretch}{1.2}
    \renewcommand{\tabcolsep}{2pt}
    \begin{tabular}{|P{9pt}|P{9pt}|P{9pt}|P{9pt}|P{9pt}|P{9pt}|P{9pt}|P{9pt}|}
        \hline
        $\vector_1$ & 0 & \cellcolor{blue!10}$\vector_3$ & 0 & $\vector_5$ & 0 & \cellcolor{blue!10}$\vector_7$ & 0 \\ \hline
        $\vector_2$ & 0 & \cellcolor{blue!10}$\vector_4$ & 0 & $\vector_6$ & 0 & \cellcolor{blue!10}$\vector_8$ & 0 \\ \hline
        $\vector_3$ & 0 & \cellcolor{blue!10}$\vector_5$ & 0 & $\vector_7$ & 0 & \cellcolor{blue!10}$\vector_1$ & 0 \\ \hline
        $\vector_4$ & 0 & \cellcolor{blue!10}$\vector_6$ & 0 & $\vector_8$ & 0 & \cellcolor{blue!10}$\vector_2$ & 0 \\ \hline
        $\vector_5$ & 0 & \cellcolor{blue!10}$\vector_7$ & 0 & $\vector_1$ & 0 & \cellcolor{blue!10}$\vector_3$ & 0 \\ \hline
        $\vector_6$ & 0 & \cellcolor{blue!10}$\vector_8$ & 0 & $\vector_2$ & 0 & \cellcolor{blue!10}$\vector_4$ & 0 \\ \hline
        $\vector_7$ & 0 & \cellcolor{blue!10}$\vector_1$ & 0 & $\vector_3$ & 0 & \cellcolor{blue!10}$\vector_5$ & 0 \\ \hline
        $\vector_8$ & 0 & \cellcolor{blue!10}$\vector_2$ & 0 & $\vector_4$ & 0 & \cellcolor{blue!10}$\vector_6$ & 0 \\ \hline
    \end{tabular}
    &
    \renewcommand{\arraystretch}{1.2}
    \renewcommand{\tabcolsep}{2pt}
    \begin{tabular}{|P{9pt}|P{9pt}|P{9pt}|P{9pt}|P{9pt}|P{9pt}|P{9pt}|P{9pt}|}
        \hline
        $\vector_1$ & \cellcolor{blue!10}$\vector_2$ & $\vector_3$ & \cellcolor{blue!10}$\vector_4$ & $\vector_5$ & \cellcolor{blue!10}$\vector_6$ & $\vector_7$ & \cellcolor{blue!10}$\vector_8$ \\ \hline
        $\vector_2$ & \cellcolor{blue!10}$\vector_3$ & $\vector_4$ & \cellcolor{blue!10}$\vector_5$ & $\vector_6$ & \cellcolor{blue!10}$\vector_7$ & $\vector_8$ & \cellcolor{blue!10}$\vector_1$ \\ \hline
        $\vector_3$ & \cellcolor{blue!10}$\vector_4$ & $\vector_5$ & \cellcolor{blue!10}$\vector_6$ & $\vector_7$ & \cellcolor{blue!10}$\vector_8$ & $\vector_1$ & \cellcolor{blue!10}$\vector_2$ \\ \hline
        $\vector_4$ & \cellcolor{blue!10}$\vector_5$ & $\vector_6$ & \cellcolor{blue!10}$\vector_7$ & $\vector_8$ & \cellcolor{blue!10}$\vector_1$ & $\vector_2$ & \cellcolor{blue!10}$\vector_3$ \\ \hline
        $\vector_5$ & \cellcolor{blue!10}$\vector_6$ & $\vector_7$ & \cellcolor{blue!10}$\vector_8$ & $\vector_1$ & \cellcolor{blue!10}$\vector_2$ & $\vector_3$ & \cellcolor{blue!10}$\vector_4$ \\ \hline
        $\vector_6$ & \cellcolor{blue!10}$\vector_7$ & $\vector_8$ & \cellcolor{blue!10}$\vector_1$ & $\vector_2$ & \cellcolor{blue!10}$\vector_3$ & $\vector_4$ & \cellcolor{blue!10}$\vector_5$ \\ \hline
        $\vector_7$ & \cellcolor{blue!10}$\vector_8$ & $\vector_1$ & \cellcolor{blue!10}$\vector_2$ & $\vector_3$ & \cellcolor{blue!10}$\vector_4$ & $\vector_5$ & \cellcolor{blue!10}$\vector_6$ \\ \hline
        $\vector_8$ & \cellcolor{blue!10}$\vector_1$ & $\vector_2$ & \cellcolor{blue!10}$\vector_3$ & $\vector_4$ & \cellcolor{blue!10}$\vector_5$ & $\vector_6$ & \cellcolor{blue!10}$\vector_7$ \\ \hline
    \end{tabular}
    &
    \renewcommand{\arraystretch}{1.2}
    \renewcommand{\tabcolsep}{2pt}
    \begin{tabular}{|P{9pt}|P{9pt}|P{9pt}|P{9pt}|P{9pt}|P{9pt}|P{9pt}|P{9pt}|}
        \hline
        \cellcolor{blue!10}$\vector_1$ & \cellcolor{blue!10}$\vector_2$ & \cellcolor{blue!10}$\vector_3$ & \cellcolor{blue!10}$\vector_4$ & \cellcolor{blue!10}$\vector_5$ & \cellcolor{blue!10}$\vector_6$ & \cellcolor{blue!10}$\vector_7$ & \cellcolor{blue!10}$\vector_8$ \\ \hline
        0 & 0 & 0 & 0 & 0 & 0 & 0 & 0 \\ \hline
        0 & 0 & 0 & 0 & 0 & 0 & 0 & 0 \\ \hline
        0 & 0 & 0 & 0 & 0 & 0 & 0 & 0 \\ \hline
        0 & 0 & 0 & 0 & 0 & 0 & 0 & 0 \\ \hline
        0 & 0 & 0 & 0 & 0 & 0 & 0 & 0 \\ \hline
        0 & 0 & 0 & 0 & 0 & 0 & 0 & 0 \\ \hline
        0 & 0 & 0 & 0 & 0 & 0 & 0 & 0 \\ \hline
    \end{tabular}
    \vspace{4pt}
    \\
    Starting configuration &
    Left rotation by $\vectorLength(\vectorLength - 1)/2$ &
    Left rotation by $\vectorLength(\vectorLength - 1)/4$ &
    Left rotation by $\vectorLength(\vectorLength - 1)/8$ &
    Mask all but the first row
    \vspace{-2pt}
    \\
    \colorbox{blue!10}{$v$} &
    $v \, +$ \colorbox{blue!10}{$v \ll 28$} &
    $v \, +$ \colorbox{blue!10}{$v \ll 14$} &
    $v \, +$ \colorbox{blue!10}{$v \ll 7$} &
    \MyColorBox[blue!10]{$v \cdot (1^\vectorLength \parallel 0^{\vectorLength (\vectorLength - 1)})$}
    \end{tabular}
    }
    \caption{Transposing a vector of length $\vectorLength=8$ from column to row (\transC).}
    \label{fig:transpositionExample}
\end{figure*}

We provide background information and notation regarding CKKS, along with an overview of the homomorphic functionalities upon which our design is constructed.

\subsection{CKKS Scheme}

CKKS~\cite{cheon2017homomorphic} is a fully homomorphic encryption scheme based on the RLWE problem.
It works with residual polynomial rings of the form $R_q = \Z_q[x] / (x^n + 1)$, where the ring dimension $n$ is a power of two.
Messages from $\C^{n/2}$ are encoded into plaintexts, which can embed vectors of up to $n / 2$ slots.
The scheme operates with floating-point values.
The encryption and homomorphic operations (see below) introduce noise on the underlying plaintexts, which is blended with the noise inherent in floating-point arithmetic, making the scheme intrinsically approximate.

CKKS natively supports three homomorphic operations on ciphertexts:
\begin{itemize}
    \item addition $(X+Y)$ corresponds to the component-wise addition of the underlying plaintext vectors;
    \item multiplication $(X \cdot Y)$ corresponds to the component-wise multiplication of the underlying plaintext vectors;
    \item rotation $(X \ll k)$ and $(X \gg k)$ correspond to the left and right rotations of the underlying plaintext vector by a plaintext index $k$.
\end{itemize}
The component-wise operations allow processing many inputs concurrently, which makes this encryption scheme suitable for SIMD.
Moreover, additions and multiplications can also be performed between a ciphertext and a plaintext.
Among all these operations, ciphertext-ciphertext multiplications and rotations are computationally the most expensive.

\subsection{Evaluating Non-Polynomial Functions}

By combining additions and multiplications it is possible to evaluate any polynomial under encryption.
Evaluating non-polynomial functions, such as the comparison operation, is not trivial under CKKS.
The usual solution consists of approximating the function by a polynomial.
However, a good approximation usually requires a high-degree polynomial, which leads to a deep multiplicative circuit to be evaluated and high computational cost.
Thus, this paper focuses on designing algorithms in which the number of non-polynomial evaluations is minimized.

In our design, we use two non-polynomial functions: the comparison function and the indicator function, defined as
\begin{equation}
\label{eq:cmp}
\cmp(x,y) =
\begin{cases*}
    1 & if $x > y$ \\
    0.5 & if $x = y$ \\
    0 & if $x < y$
\end{cases*},
\end{equation}
\begin{equation}
\label{eq:ind}
\ind_A(x) =
\begin{cases*}
    1 & if $x \in A$ \\
    0 & if $x \notin A$
\end{cases*}.
\end{equation}

\noindent We approximate both using Chebyshev polynomials, which assure uniform convergence to the original function.
The evaluation of the polynomials is then performed using the Paterson-Stockmeyer algorithm adapted to Chebyshev basis~\cite{CCS19}, which requires only $O(\sqrt \approximationDegree)$ homomorphic multiplications to evaluate a degree-$\approximationDegree$ polynomial.
We denote an approximation of degree $\approximationDegree$ of these functions with $\cmp(\,\cdot\,, \,\cdot\,; \approximationDegree)$ and $\ind_A(\,\cdot\,; \approximationDegree)$, respectively.
In terms of multiplicative depth, each approximation requires around $\log(\approximationDegree)$ levels.

For assessing our solution against related work, we also implement the comparison function by Cheon et al.~\cite{cheon2020efficient}, which is used in the work of Hong et al.~\cite{hong2021efficient}.
Their implementation is based on the composition of two polynomials
\begin{align*}
    f(x) &= (35 x - 35 x^3 + 21 x^5 - 5 x^7) / 2^4 \\
    g(x) &= (4589 x - 16577 x^3 + 25614 x^5- 12860 x^7) / 2^{10} \enspace .
\end{align*}
In particular, by combining $d_f$ times $f$ and $d_g$ times $g$, one can get an arbitrarily close approximation of the compare function $\cmp(x) = (f^{d_f}(g^{d_g}(x-y)) + 1) / 2$.
While the indicator function for any interval $A = [a,b]$ can be computed as $\ind_{[a,b]}(x) = \cmp(x, a) (1 - \cmp(x, b))$.

Chebyshev usually leads to approximations with low multiplicative depth, while the $f,g$ approach leads to approximations with low evaluation runtime.
In general, for a comparison function, the higher the degree of the approximation the better it can correctly compare values that are close to each other.
For a discussion on this topic and for different implementations of the comparison function, we refer to \cite{lee2021minimax}.

\subsection{Matrix Encoding and Operations}

Our approach relies on matrices for intermediate computations.
To encode a matrix into a ciphertext we adopt the row-by-row approach~\cite{kim2018logistic}, which consists of concatenating each row into a single vector and then encrypting it.
For a square matrix of size $\vectorLength$, we have the requirement that $\vectorLength^2 \le n/2$, where $n$ is the ring dimension, otherwise multiple ciphertexts are needed to store the entire matrix.

We describe some basic functionalities useful for working with encrypted matrices.
These functionalities will be the building blocks of our approach.
Given a matrix encoded into a ciphertext $X$:
\begin{itemize}
    \item $\maskR(X, k)$ extracts row $k$ by masking everything else, i.e., setting everything else to zero;
    \item $\sumR(X)$ sums all the rows together component-wise and stores the result in the first row;
    \item $\replR(X)$ assumes a matrix with only the first row non-zero and replicates it all over by copying its values into the other rows;
    \item $\transR(X)$ assumes a square matrix with only the first row non-zero and transposes it (i.e., moving it into the first column).
\end{itemize}
Similarly for the columns, we have $\maskC$, $\sumC$, $\replC$, $\transC$.
Masking works by multiplying the encrypted matrix by an appropriate plaintext bit mask.
For sum and replication there are well-known algorithms in the literature that work recursively, and only require $\log(\vectorLength)$ rotations, where $\vectorLength$ is the number of rows/columns of the matrix~\cite{halevi2014algorithms}.
For these to work, $\vectorLength$ must be a power of two, thus the matrix might require padding.
We provide their pseudocode in \Cref{app:rec_matrix_ops}.
As for transposition, to the best of our knowledge no algorithm that works in $\log(\vectorLength)$ is available in the literature.
Hence, we propose \Cref{alg:transR} and \Cref{alg:transC}, which can be of independent interest.
\Cref{fig:transpositionExample} shows an example of the \transC \, algorithm for a generic vector of length $\vectorLength = 8$.

\begin{algorithm}[htb]
\caption{\transR}
\label{alg:transR}
\begin{algorithmic}[1]

\Require $X$ encryption of a vector $x$ encoded as a row.

\Ensure $X$ encryption of the vector $x$ encoded as a column.

\For{$i = 1, \dots, \ceil{\log{\vectorLength}}$}
    \State $X \gets X + (X \gg \vectorLength(\vectorLength-1) / 2^i)$
\EndFor
\State $X \gets \maskC(X,0)$
\State \Return $X$
\end{algorithmic}
\end{algorithm}

\begin{algorithm}[htb]
\caption{\transC}
\label{alg:transC}
\begin{algorithmic}[1]

\Require $X$ encryption of a vector $x$ encoded as a column.

\Ensure $X$ encryption of the vector $x$ encoded as a row.

\For{$i = 1, \dots, \ceil{\log{\vectorLength}}$}
    \State $X \gets X + (X \ll \vectorLength(\vectorLength-1) / 2^i)$
\EndFor
\State $X \gets \maskR(X,0)$
\State \Return $X$
\end{algorithmic}
\end{algorithm}


\section{Our Design for Ranking, Order statistics, and Sorting}
\label{sec:main-design}

The core idea of our design is to manipulate the encrypted vector in such a way that \textbf{only a single evaluation of the comparison function is needed to compare all values}.
For instance, given a vector $\vector = (\vector_1, \vector_2, \vector_3)$, we produce
$$\vector_R = (\vector_1, \vector_2, \vector_3, \vector_1, \vector_2, \vector_3, \vector_1, \vector_2, \vector_3),$$
$$\vector_C = (\vector_1, \vector_1, \vector_1, \vector_2, \vector_2, \vector_2, \vector_3, \vector_3, \vector_3).$$
The comparison $\cmp(\vector_R, \vector_C)$ contains information about $\vector_i < \vector_j$ for all pairs $(\vector_i, \vector_j)$.
It is easier to visualize this by seeing $\vector_R, \vector_C$ as square matrices that have been encoded row-by-row into vectors:
$$
\vector_R =
\begin{pmatrix}
\vector_1 & \vector_2 & \vector_3 \\
\vector_1 & \vector_2 & \vector_3 \\
\vector_1 & \vector_2 & \vector_3 \\
\end{pmatrix},
\qquad
\vector_C =
\begin{pmatrix}
\vector_1 & \vector_1 & \vector_1 \\
\vector_2 & \vector_2 & \vector_2 \\
\vector_3 & \vector_3 & \vector_3 \\
\end{pmatrix}.
$$
Turning $\vector$ into $\vector_R$ and $\vector_C$ homomorphically can be done by composing the matrix operations mentioned above, as will be described in detail later.
Note that we are implicitly assuming we can fit $\vectorLength^2$ elements in a ciphertext, where the vector length $\vectorLength$ is 3 in the example above.
If this assumption does not hold, we require multiple ciphertexts, which we discuss in \Cref{sec:multiple_ciphertexts}.

\subsection{Ranking}

Ranking associates the elements of a vector to their rank, that is the position they would have if the vector was sorted, starting from $1$.
In case of ties, we adopt fractional ranking, for which ties receive a rank equal to the average of the ranks they span.
For instance, the ranking of $(50,10,20,20,40)$ is $(5,1,2.5,2.5,4)$.

\begin{figure}[!htb]
    \centering
    \begin{tikzpicture}[node distance=85pt]

    \tikzstyle{every node}=[font=\footnotesize]

    \node (table1) {
    \renewcommand{\tabcolsep}{\tabcolsepValue}
    \renewcommand{\arraystretch}{\arraystretchValue}
    \begin{tabular}{|P{10pt}|P{10pt}|P{10pt}|P{10pt}|}
        \hline
        20 & 30 & 10 & 40 \\ \hline
        0 & 0 & 0 & 0 \\ \hline
        0 & 0 & 0 & 0 \\ \hline
        0 & 0 & 0 & 0 \\ \hline
    \end{tabular}
    };

    \node[below of=table1] (table2) {
    \renewcommand{\tabcolsep}{\tabcolsepValue}
    \renewcommand{\arraystretch}{\arraystretchValue}
    \begin{tabular}{|P{10pt}|P{10pt}|P{10pt}|P{10pt}|}
        \hline
        20 & 30 & 10 & 40 \\ \hline
        20 & 30 & 10 & 40 \\ \hline
        20 & 30 & 10 & 40 \\ \hline
        20 & 30 & 10 & 40 \\ \hline
    \end{tabular}
    };

    \node[left of=table1] (table3) {
    \renewcommand{\tabcolsep}{\tabcolsepValue}
    \renewcommand{\arraystretch}{\arraystretchValue}
    \begin{tabular}{|P{10pt}|P{10pt}|P{10pt}|P{10pt}|}
        \hline
        20 & 0 & 0 & 0 \\ \hline
        30 & 0 & 0 & 0 \\ \hline
        10 & 0 & 0 & 0 \\ \hline
        40 & 0 & 0 & 0 \\ \hline
    \end{tabular}
    };

    \node[below of=table3] (table4) {
    \renewcommand{\tabcolsep}{\tabcolsepValue}
    \renewcommand{\arraystretch}{\arraystretchValue}
    \begin{tabular}{|P{10pt}|P{10pt}|P{10pt}|P{10pt}|}
        \hline
        20 & 20 & 20 & 20 \\ \hline
        30 & 30 & 30 & 30 \\ \hline
        10 & 10 & 10 & 10 \\ \hline
        40 & 40 & 40 & 40 \\ \hline
    \end{tabular}
    };

    \node[draw, rounded corners=2pt, below=20pt of table4, text width=30pt, align=center] (comp) {
        $\cmp$ $o_1 > o_2$
    };
    
    \node[below=20pt of comp] (table5) {
    \renewcommand{\tabcolsep}{\tabcolsepValue}
    \renewcommand{\arraystretch}{\arraystretchValue}
    \begin{tabular}{|P{10pt}|P{10pt}|P{10pt}|P{10pt}|}
        \hline
        0.5 & 1 & 0 & 1 \\ \hline
        0 & 0.5 & 0 & 1 \\ \hline
        1 & 1 & 0.5 & 1 \\ \hline
        0 & 0 & 0 & 0.5 \\ \hline
    \end{tabular}
    };

    \node[right of=table5] (table6) {
    \renewcommand{\tabcolsep}{\tabcolsepValue}
    \renewcommand{\arraystretch}{\arraystretchValue}
    \begin{tabular}{|P{10pt}|P{10pt}|P{10pt}|P{10pt}|}
        \hline
        1.5 & 2.5 & 0.5 & 3.5 \\ \hline
        0 & 0 & 0 & 0 \\ \hline
        0 & 0 & 0 & 0 \\ \hline
        0 & 0 & 0 & 0 \\ \hline
    \end{tabular}
    };

    \node[right of=table6] (table7) {
    \renewcommand{\tabcolsep}{\tabcolsepValue}
    \renewcommand{\arraystretch}{\arraystretchValue}
    \begin{tabular}{|P{10pt}|P{10pt}|P{10pt}|P{10pt}|}
        \hline
        2 & 3 & 1 & 4 \\ \hline
        0 & 0 & 0 & 0 \\ \hline
        0 & 0 & 0 & 0 \\ \hline
        0 & 0 & 0 & 0 \\ \hline
    \end{tabular}
    };

    \node[above=-2pt of table1] {\textbf{Input}};
    \node[above=-2pt of table7] {\textbf{Output}};

    \draw[->] (table1) -- (table2) node[midway, right] {\replR};
    \draw[->, shorten <=-4pt] (table1) -- (table3) node[pos=0.38, above] {\transR};
    \draw[->] (table3) -- (table4) node[midway, right] {\replC};
    \draw[->, shorten >=3pt] (table2) |- (comp) node[near start, right] {$o_1$};
    \draw[->, shorten >=3pt] (table4) -- (comp) node[midway, right] {$o_2$};
    \draw[->, shorten <=3pt] (comp) -- (table5) node[midway, above] {};
    \draw[->, shorten >=-4pt] (table5) -- (table6) node[pos=0.62, above] {\sumR};
    \draw[->, shorten >=-4pt] (table6) -- (table7) node[midway, above] {$+ 0.5$};

    \end{tikzpicture}
    
    \caption{Schematic example of ranking a 4-element vector.}
    \label{fig:schematicExampleRanking}
\end{figure}
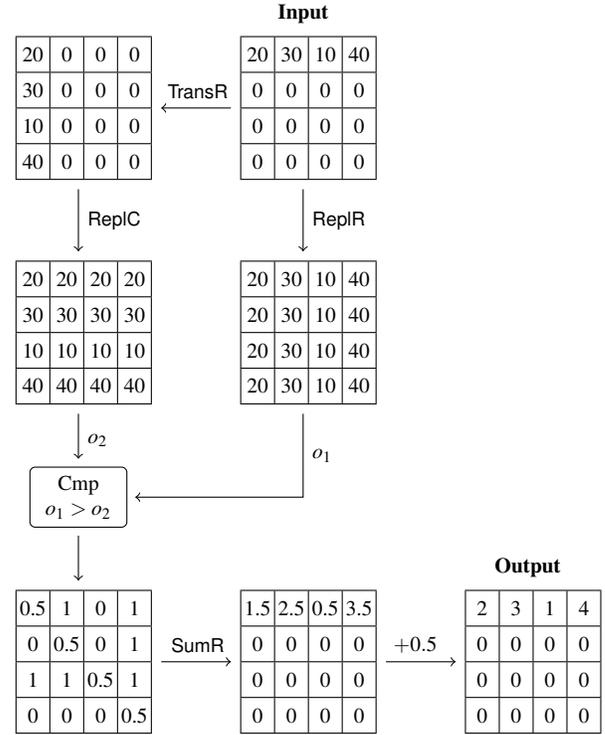

Given an input vector $\vector$ encrypted as $\encVector$, we think of it as the first row of a null matrix.
The encoding $\vector_R$ is produced by simply applying \replR, while $\vector_C$ is produced by first transposing the initial vector to a column with \transR \, and then replicating it with \replC.
The component-wise comparison of $\vector_R > \vector_C$ is ideally a matrix with values in $\{0, 0.5, 1\}$, whose each column $j$ contains information about the position of $\vector_j$ in the sorted array:
\begin{itemize}
    \item a number of ones equal to the number of elements smaller than $\vector_j$, and
    \item a number of 0.5 equal to the number of elements equal to $\vector_j$.
\end{itemize}
Thus, summing the elements in the column (and an additional 0.5) gives the fractional rank of $\vector_j$.
A pseudocode is provided in \Cref{alg:ranking}, while a schematic with an example can be found in \Cref{fig:schematicExampleRanking}.

\begin{algorithm}[!htb]
\caption{\rank}
\label{alg:ranking}
\begin{algorithmic}[1]
\Require $\encVector$ encryption of $\vector = (\vector_1, \dots, \vector_\vectorLength) \in \R^\vectorLength$, approximation degree $\approximationDegree \in \N$.
\Ensure $\encRanking$ encryption of a vector in $\R^\vectorLength$ representing the (fractional) ranking of $\vector$.
\State $\encVector_R \gets \replR(\encVector)$
\State $\encVector_C \gets \replC(\transR(\encVector))$
\State $\encComparison \gets \cmp(\encVector_R, \encVector_C; \approximationDegree)$
\State $\encRanking \gets \sumR(\encComparison) + (0.5, \dots, 0.5)$
\State \Return $\encRanking$
\end{algorithmic}
\end{algorithm}

We assume the vector size is a power of 2 to work with recursive operations on matrices. If it is not, we can pad it to the next power of 2 and perform an additional masking after the comparison to remove the excess information.

The cost of \Cref{alg:ranking} is $4 \ceil{\log{\vectorLength}}$ rotations, which can be reduced to $3 \ceil{\log{\vectorLength}}$ by parallelizing \replR \, and \replC, and $\sqrt{\approximationDegree}$ ciphertext-ciphertext multiplications, with a multiplicative depth of $\mytilde\ceil{\log{\approximationDegree}}$.

\paragraph{Correctness Proof}
Assuming the correctness of the building blocks and the ideal functionality of the algorithm, namely no noise from the scheme and no approximation error for the comparison function, we prove that the output $\encRanking$ produced by \Cref{alg:ranking} on input the encryption of a vector $\vector = (\vector_1, \dots, \vector_\vectorLength)$ is actually the encryption of the fractional ranking of $\vector$.
Let $\ranking = (\ranking_1, \dots, \ranking_\vectorLength)$ be the decryption of $\encRanking$.
Let $j \in \{1,\dots,\vectorLength\}$, we prove that $r_j$ is the rank of $v_j$.
Let $\vector_R$ and $\vector_C$ be the decryption as matrices of $\encVector_R$ and $\encVector_C$, respectively.
By the correctness of \replR, $\vector_{R;i,j} = v_j$ for all $i \in \{1, \dots, \vectorLength\}$, where $\vector_{R;i,j}$ is the element at row $i$ and column $j$ of $\vector_{R}$.
Similarly, by the correctness of \replC \, and \transR, $\vector_{C;i,j} = v_i$ for all $i \in \{1, \dots, \vectorLength\}$.
Let $\comparison$ be the decryption as matrix of $\encComparison$, then $\comparison_{i,j} = \cmp(\vector_j, \vector_i)$, where $\cmp$ is defined in Equation~\ref{eq:cmp}.
Then $r_j = \sum_{i = 1}^{\vectorLength}{\comparison_{i,j}} + 0.5$.
We split the sum over the following partition of $\{1, \dots, \vectorLength\}$:
\begin{align*}
    \partitionLower_j &:= \{ i \in \{1,\dots,\vectorLength\} : v_i < v_j\} \\
    \partitionEqual_j &:= \{ i \in \{1,\dots,\vectorLength\} : v_i = v_j\} \\
    \partitionGreater_j &:= \{ i \in \{1,\dots,\vectorLength\} : v_i > v_j\}
\end{align*}
and get
\begin{align*}
    r_j &= \sum_{\partitionLower_j}{\comparison_{i,j}} + \sum_{\partitionEqual_j}{\comparison_{i,j}} + \sum_{\partitionGreater_j}{\comparison_{i,j}} + 0.5 \\
    &= \sum_{\partitionLower_j}{1} + \sum_{\partitionEqual_j}{0.5} + \sum_{\partitionGreater_j}{0} + 0.5 \\
    &= |\partitionLower_j| + 0.5 \cdot |\partitionEqual_j| + 0.5 .
\end{align*}
The elements equal to $\vector_j$ span a rank from $|\partitionLower_j| + 1$ to $|\partitionLower_j| + |\partitionEqual_j|$, thus the fractional rank of $\vector_j$ is their average, namely
\begin{align*}
    \frac{1}{|\partitionEqual_j|} \sum_{k = 1}^{|\partitionEqual_j|}{\left(|\partitionLower_j| + k\right)}
    &= \frac{1}{|\partitionEqual_j|} \left(|\partitionEqual_j| \cdot |\partitionLower_j| + 0.5 \cdot |\partitionEqual_j| \cdot \left(|\partitionEqual_j| + 1\right)\right) \\
    &= |\partitionLower_j| + 0.5 \cdot \left(|\partitionEqual_j| + 1\right) = r_j .\qquad\quad \qed
\end{align*}

\subsection{Order Statistics}

The $k$-th order statistic (or k-statistic) of a vector is its $k$-th smallest value, that is the value of rank $k$ if such rank exists.
A value of given rank might not exist if there are ties in the vector. Here, we will show how to work around this issue in the specific case of the first and last order statistics (i.e., min and max). While we will provide a general-purpose solution in \Cref{sec:handling-ties}.

We can determine the $k$-th order statistic of a vector $\vector$ by first computing its ranking and then applying to it an indicator function ``around $k$'', namely for the interval $[k-0.5,k+0.5]$.
It will output a bit mask whose ones correspond to the positions of the elements with rank $k$, if they exist (see \Cref{alg:order_statistic}).
\begin{algorithm}[!htb]
\caption{Order Statistic}
\label{alg:order_statistic}
\begin{algorithmic}[1]
\Require $\encVector$ encryption of $\vector = (\vector_1, \dots, \vector_\vectorLength) \in \R^\vectorLength$, approximation degrees $\approximationDegree_C, \approximationDegree_I \in \N$, index $k \in \{1, \dots, \vectorLength\}$.
\Ensure $\encOrderStat$ encryption of a Boolean vector in $\{0,1\}^\vectorLength$ that has value 1 in position $i$ if and only if $\vector_i$ has rank $k$.
\State $\encRanking \gets \rank(\encVector; \approximationDegree_C)$
\State $\encOrderStat \gets \ind_k(\encRanking; \approximationDegree_I)$
\State \Return $\encOrderStat$
\end{algorithmic}
\end{algorithm}
One can then retrieve the actual value of the statistic by computing the inner product $\innerproduct{\encOrderStat}{\encVector} = \sumC(\encOrderStat \cdot \encVector)$ and dividing it by the L1 norm of the mask $\sumC(\encOrderStat)$.
This can be done under encryption by approximating $1/x$ in the range $[0.5, \vectorLength + 0.5]$, or by using Goldschmidt's division algorithm~\cite{cheon2019numerical}.
The outcome will be zero if no element of rank $k$ exists.

\paragraph{Correctness Proof}
Assuming the correctness of the building blocks and \Cref{alg:ranking}, and the ideal functionality of the algorithm, we prove that \Cref{alg:order_statistic} is correct.
The input is the encryption of a vector $\vector = (\vector_1, \dots, \vector_\vectorLength)$, together with an index $k \in \N$.
Let $\ranking, \orderStat$ be the decryption of $\encRanking, \encOrderStat$ respectively.
For $i \in \{1, \dots, \vectorLength\}$,
$$\,\,\,\,\quad\qquad o_i = \ind_k(\ranking_i) = \begin{cases*}
    1 & if $\rank(\vector_i) = k$ \\
    0 & if $\rank(\vector_i) \ne k$
\end{cases*} . \qquad\quad\,\,\,\, \qed$$

\paragraph{Min and Max}
We can ensure that we can always compute minimum and maximum (first and last order statistic) by employing a slightly different definition of the comparison function.
By using
\begin{equation*}
\cmp_G(x,y) =
\begin{cases*}
    1 & if $x > y$ \\
    0 & if $x \le y$
\end{cases*}   
\end{equation*}
all the minimal elements are assigned to the first rank, thus the minimum can be retrieved with $\ind_1$.
Similarly, by using
\begin{equation*}
\cmp_{GE}(x,y) =
\begin{cases*}
    1 & if $x \ge y$ \\
    0 & if $x < y$
\end{cases*}   
\end{equation*}
all the maximal elements are assigned to the last rank, thus the maximum can be retrieved with $\ind_\vectorLength$.

\subsection{Sorting}

We sort a vector by extracting all its order statistics simultaneously, parallelizing the strategy presented in \Cref{alg:order_statistic}.
For now, we assume that all elements in the vector are distinct,
ensuring that there is exactly one element for each rank $k \in \{1, \dots, \vectorLength\}$,
and allowing us to extract any order statistic.
We will show how to remove this assumption in \Cref{sec:handling-ties}.

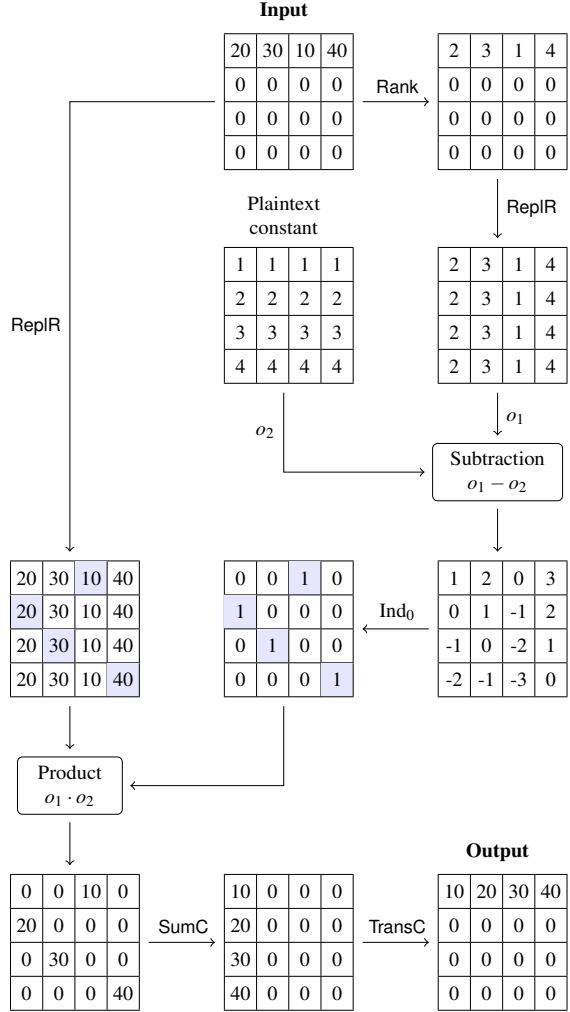
\begin{figure}[!t]
    \centering
    \resizebox{!}{388pt}{%
    \begin{tikzpicture}[node distance=85pt]

    \tikzstyle{every node}=[font=\footnotesize]
    
    \node (table1) {
    \renewcommand{\tabcolsep}{\tabcolsepValue}
    \renewcommand{\arraystretch}{\arraystretchValue}
    \begin{tabular}{|P{10pt}|P{10pt}|P{10pt}|P{10pt}|}
        \hline
        20 & 30 & 10 & 40 \\ \hline
        0 & 0 & 0 & 0 \\ \hline
        0 & 0 & 0 & 0 \\ \hline
        0 & 0 & 0 & 0 \\ \hline
    \end{tabular}
    };

    \node[right of=table1] (table2) {
    \renewcommand{\tabcolsep}{\tabcolsepValue}
    \renewcommand{\arraystretch}{\arraystretchValue}
    \begin{tabular}{|P{10pt}|P{10pt}|P{10pt}|P{10pt}|}
        \hline
        2 & 3 & 1 & 4 \\ \hline
        0 & 0 & 0 & 0 \\ \hline
        0 & 0 & 0 & 0 \\ \hline
        0 & 0 & 0 & 0 \\ \hline
    \end{tabular}
    };

    \node[below of=table2] (table3) {
    \renewcommand{\tabcolsep}{\tabcolsepValue}
    \renewcommand{\arraystretch}{\arraystretchValue}
    \begin{tabular}{|P{10pt}|P{10pt}|P{10pt}|P{10pt}|}
        \hline
        2 & 3 & 1 & 4 \\ \hline
        2 & 3 & 1 & 4 \\ \hline
        2 & 3 & 1 & 4 \\ \hline
        2 & 3 & 1 & 4 \\ \hline
    \end{tabular}
    };

    \node[below of=table1] (table4) {
    \renewcommand{\tabcolsep}{\tabcolsepValue}
    \renewcommand{\arraystretch}{\arraystretchValue}
    \begin{tabular}{|P{10pt}|P{10pt}|P{10pt}|P{10pt}|}
        \hline
        1 & 1 & 1 & 1 \\ \hline
        2 & 2 & 2 & 2 \\ \hline
        3 & 3 & 3 & 3 \\ \hline
        4 & 4 & 4 & 4 \\ \hline
    \end{tabular}
    };

    \node[draw, rounded corners=2pt, below=20pt of table3, text width=45pt, align=center] (minus) {
    Subtraction $o_1 - o_2$
    };

    \node[below=20pt of minus] (table5) {
    \renewcommand{\tabcolsep}{\tabcolsepValue}
    \renewcommand{\arraystretch}{\arraystretchValue}
    \begin{tabular}{|P{10pt}|P{10pt}|P{10pt}|P{10pt}|}
        \hline
         1 &  2 &  0 &  3 \\ \hline
         0 &  1 & -1 &  2 \\ \hline
        -1 &  0 & -2 &  1 \\ \hline
        -2 & -1 & -3 &  0 \\ \hline
    \end{tabular}
    };

    \node[left of=table5] (table6) {
    \renewcommand{\tabcolsep}{\tabcolsepValue}
    \renewcommand{\arraystretch}{\arraystretchValue}
    \begin{tabular}{|P{10pt}|P{10pt}|P{10pt}|P{10pt}|}
        \hline
        0 & 0 & \cellcolor{blue!10}1 & 0 \\ \hline
        \cellcolor{blue!10}1 & 0 & 0 & 0 \\ \hline
        0 & \cellcolor{blue!10}1 & 0 & 0 \\ \hline
        0 & 0 & 0 & \cellcolor{blue!10}1 \\ \hline
    \end{tabular}
    };

    \node[left of=table6] (table7) {
    \renewcommand{\tabcolsep}{\tabcolsepValue}
    \renewcommand{\arraystretch}{\arraystretchValue}
    \begin{tabular}{|P{10pt}|P{10pt}|P{10pt}|P{10pt}|}
        \hline
        20 & 30 & \cellcolor{blue!10}10 & 40 \\ \hline
        \cellcolor{blue!10}20 & 30 & 10 & 40 \\ \hline
        20 & \cellcolor{blue!10}30 & 10 & 40 \\ \hline
        20 & 30 & 10 & \cellcolor{blue!10}40 \\ \hline
    \end{tabular}
    };

    \node[draw, rounded corners=2pt, below=20pt of table7, text width=35pt, align=center] (mult) {
    Product $o_1 \cdot o_2$
    };

    \node[below=20pt of mult] (table8) {
    \renewcommand{\tabcolsep}{\tabcolsepValue}
    \renewcommand{\arraystretch}{\arraystretchValue}
    \begin{tabular}{|P{10pt}|P{10pt}|P{10pt}|P{10pt}|}
        \hline
         0 &  0 & 10 &  0 \\ \hline
        20 &  0 &  0 &  0 \\ \hline
         0 & 30 &  0 &  0 \\ \hline
         0 &  0 &  0 & 40 \\ \hline
    \end{tabular}
    };

    \node[right of=table8] (table9) {
    \renewcommand{\tabcolsep}{\tabcolsepValue}
    \renewcommand{\arraystretch}{\arraystretchValue}
    \begin{tabular}{|P{10pt}|P{10pt}|P{10pt}|P{10pt}|}
        \hline
        10 & 0 & 0 & 0 \\ \hline
        20 & 0 & 0 & 0 \\ \hline
        30 & 0 & 0 & 0 \\ \hline
        40 & 0 & 0 & 0 \\ \hline
    \end{tabular}
    };

    \node[right of=table9] (table10) {
    \renewcommand{\tabcolsep}{\tabcolsepValue}
    \renewcommand{\arraystretch}{\arraystretchValue}
    \begin{tabular}{|P{10pt}|P{10pt}|P{10pt}|P{10pt}|}
        \hline
        10 & 20 & 30 & 40 \\ \hline
        0 & 0 & 0 & 0 \\ \hline
        0 & 0 & 0 & 0 \\ \hline
        0 & 0 & 0 & 0 \\ \hline
    \end{tabular}
    };

    \node[above=-2pt of table1] {\textbf{Input}};
    \node[above=-2pt of table10] {\textbf{Output}};
    \node[above=-2pt of table4, text width=35pt, align=center] {Plaintext constant};

    \draw[->, shorten >=-4pt] (table1) -- (table2) node[pos=0.62, above] {\rank};
    \draw[->] (table2) -- (table3) node[midway, right] {\replR};
    \draw[->, shorten >=3pt] (table3) -- (minus) node[midway, right] {$o_1$};
    \draw[->, shorten >=3pt] (table4) |- (minus) node[near start, left] {$o_2$};
    \draw[->, shorten <=3pt] (minus) -- (table5) node[midway, right] {};
    \draw[->, shorten <=-4pt] (table5) -- (table6) node[pos=0.38, above] {$\ind_0$};
    \draw[->, shorten <=-4pt] (table1) -| (table7) node[near end, left] {\replR};
    \draw[->, shorten >=3pt] (table7) -- (mult) node[midway, right] {};
    \draw[->, shorten >=3pt] (table6) |- (mult) node[midway, right] {};
    \draw[->, shorten <=3pt] (mult) -- (table8) node[midway, right] {};
    \draw[->, shorten >=-4pt] (table8) -- (table9) node[pos=0.62, above] {\sumC};
    \draw[->, shorten >=-4pt] (table9) -- (table10) node[pos=0.62, above] {\transC};

    \end{tikzpicture}
    }
    
    \caption{Schematic example of sorting a 4-element vector.}
    \label{fig:schematicExampleSorting}
\end{figure}

To extract all the order statistics in one go, the idea is to compute the ranking of $\vector$, replicate it over the rows, and extract a different k-statistic for each row $k$.
Normally, this would require applying $\vectorLength$ different indicator functions.
To avoid this, we shift each row's domain by subtracting the entire row by $(k, \dots, k)$.
Applying an indicator function around zero then produces a one-hot encoding of the position of the k-statistic for each row $k$.
Next, we perform an inner-product with the original vector: we replicate the vector over the rows, multiply it by the previously-computed mask, and sum the result over the columns.
This way, the first element of each row $k$ contains the corresponding k-statistic of $\vector$.
The pseudocode is provided in \Cref{alg:sorting}, while a schematic with an example can be found in \Cref{fig:schematicExampleSorting}.

\begin{algorithm}[t]
\caption{Sorting}
\label{alg:sorting}
\begin{algorithmic}[1]
\Require $\encVector$ encryption of $\vector = (\vector_1, \dots, \vector_\vectorLength) \in \R^\vectorLength$ with distinct elements, approximation degrees $\approximationDegree_C, \approximationDegree_I \in \N$.
\Ensure $\encSortedVec$ encryption of the sorted form of $\vector$.
\State $\encRanking \gets \rank(\encVector; \approximationDegree_C)$
\State $\encRanking_R \gets \replR(\encRanking)$
\State $\encSortingMask \gets \ind_0(\encRanking_R - (1^\vectorLength \parallel \cdots \parallel \vectorLength^\vectorLength); \approximationDegree_I)$
\State $\encVector_R \gets \replR(\encVector)$
\State $\encSortedVec \gets \transC(\sumC(\encSortingMask \cdot \encVector_R))$
\State \Return $\encSortedVec$
\end{algorithmic}
\end{algorithm}

Note that the quantity $\replR(\encVector)$ is already computed within the ranking algorithm, thus the extra cost is given by only $3 \ceil{\log{\vectorLength}}$ rotations, the evaluation of the indicator function, and the multiplication $\encSortingMask \cdot \encVector_R$.
As an additional optimization, we can avoid the final transposition and save $\ceil{\log{\vectorLength}}$ rotations with a slight tweak in the ranking algorithm.
By inverting the order of the operands in the $\cmp$ and replacing the \sumR \, with a \sumC, the output of the ranking algorithm will be in column-form instead of row-form.
This entails swapping row- and column-operations in the sorting algorithm, and replacing $\replR(\encVector)$ with $\replC(\transR(\encVector))$, which is also a quantity computed within the ranking algorithm.
The (parallel) cost of the sorting algorithm is then:
\begin{itemize}
    \item $5 \ceil{\log{\vectorLength}}$ rotations, and
    \item $\sqrt{\approximationDegree_C} + \sqrt{\approximationDegree_I} + 1$ ciphertext-ciphertext multiplications, with a multiplicative depth of $\mytilde\ceil{\log{\approximationDegree_C}} + \ceil{\log{\approximationDegree_I}} + 1$.
\end{itemize}

\paragraph{Correctness Proof}
Assuming the correctness of the building blocks and \Cref{alg:ranking}, and the ideal functionality of the algorithm, we prove that the output $\encSortedVec$ produced by \Cref{alg:sorting} on input the encryption of a vector with distinct elements $\vector = (\vector_1, \dots, \vector_\vectorLength)$ is actually the encryption of the sorted form of $\vector$.
Let $\ranking_R, \sortingMask, \vector_R, \sortedVec$ be the decryption of $\encRanking_R, \encSortingMask, \encVector_R, \encSortedVec$ respectively.
As the elements $\vector_i$ are all distinct, the ranking of $\vector$ is a permutation of $\{1,\dots,\vectorLength\}$.
Let $i \in \{1, \dots, \vectorLength\}$, we thus have to prove that $\rank(\sortedVec_i) = i$.
By the correctness of \replR:
$$\sortingMask_{i,j} = \ind_0(\ranking_{R;i,j} - i) = \ind_0(\ranking_j - i) = \begin{cases*}
    1 & if $\ranking_j = i$ \\
    0 & if $\ranking_j \ne i$
\end{cases*}$$
and by the correctness of \sumC, \transC, and \replR, we have
$$\sortedVec_i
= \sum_{j = 1}^{\vectorLength}{\sortingMask_{i,j} \cdot \vector_{R;i,j}}
= \sum_{j = 1}^{\vectorLength}{\sortingMask_{i,j} \cdot \vector_j} .$$
Since $\ranking$ is a permutation of $\{1,\dots,\vectorLength\}$, there exists one and only one index $k$ such that $\rank(\vector_k) = i$.
Hence, $\sortedVec_i = \vector_k$ and ${\rank(\sortedVec_i) = i}$.
\qed


\section{Tie-Correction Offset}
\label{sec:handling-ties}

If two or more elements are in a \textit{tie}, namely share the same value, they receive the same rank.
As noted in \Cref{sec:main-design}, this causes the ranking function to become non-surjective, which hinders the extraction of certain order statistics and, consequently, prevents a correct sorting of the input vector.
For example, the (fractional) ranking of the input vector $\vector = [10, 20, 20, 40]$ is $\ranking = [1, 2.5, 2.5, 4]$. If we now want to extract the second or third order statistics (which should both correspond to the value $20$), we should apply an indicator around rank $2$ and $3$, respectively, which would miss the actual rank value $2.5$.
To fix this issue, we build an offset vector that redistributes the fractional ranking of all elements in a tie over the ranks they span.
In our example, the offset vector would be $\offsetVec = [0, -0.5, 0.5, 0]$, which corrects the fractional ranking to
$$\ranking + \offsetVec = [1, 2, 3, 4]$$
allowing us to correctly extract all four order statistics.
As follows, we explain how to build this tie-correction offset vector under encryption with small computational overhead.

To build this offset, we need to evaluate the equality operator ($\eq$) among all pairs of elements in the input vector.
Similarly to $\cmp$, the output of this function is a square matrix $\equality$ of size $\vectorLength \times \vectorLength$ such that $\equality_{i,j} = 1$ if $\vector_i = \vector_j$, and $0$ otherwise.
The equality can be evaluated as an indicator function around $0$, which can be done in parallel to the greater-than $\cmp$ in the ranking.
However, we note that the information needed to compute the equality matrix $\equality$ is already contained in the comparison matrix $\comparison$.
We can reuse it to compute the equality:
$$\equality = 4 \cdot \comparison \cdot (1 - \comparison)$$
mapping both zeros and ones of $\comparison$ to $0$, and the values $0.5$ to $1$.
In our implementation we mainly use the latter option, which comes with an overhead of just two multiplications.

Note that each column $j$ contains a number of ones equal to the number of elements that are in a tie with $\vector_j$.
This includes the trivial equality $\vector_j = \vector_j$ on the main diagonal.
By masking out the lower triangle of $\equality$ and summing over the rows, we count the non-trivial identities only once, namely
$$\scalingEquality_j = |\{i \in \{1, \dots, j\} : v_i = v_j \}| \enspace .$$
For instance, if the first four elements of $\vector$ are in a tie, then the corresponding values in $\scalingEquality$ are $1, 2, 3, 4$.
We can use $\scalingEquality$ to offset the ranking. But, since we are using fractional ranking, we first need to shift it by half of the tie size, that is the range the elements in the tie span.
To do this, we sum directly over the rows of $\equality$, without masking it, and get
$$\tieSize_j = |\{i \in \{1, \dots, \vectorLength\} : v_i = v_j \}| \enspace .$$
Now, the correction offset for the ranking can be computed as
$$\offsetVec_j = \scalingEquality_j - 0.5 \cdot \tieSize_j - 0.5$$
where the last $-0.5$ makes the offset start from zero, and it nicely cancels out with the $+0.5$ in the last line of the ranking algorithm.
The pseudocode to compute the tie-correction offset is presented in \Cref{alg:tie-correction-offset}.
This runs at the end of the ranking algorithm (\Cref{alg:ranking}), and the offset can just be added to the fractional ranking to make it suitable for order statistics extraction and sorting.

As a particular case, we can modify \Cref{alg:order_statistic} to compute the \textit{median} by extracting the $(\vectorLength + 1)/2$ statistic if $\vectorLength$ is odd, or both the $\vectorLength/2$ and $(\vectorLength/2)+1$ statistics if $\vectorLength$ is even.
In the latter case, an additional plaintext multiplication by $0.5$ is needed after the inner product.
In a similar way, one can compute any percentile of the given vector.

\begin{algorithm}[h]
\caption{Tie-Correction Offset}
\label{alg:tie-correction-offset}
\begin{algorithmic}[1]
\Require $\encVector$ encryption of $\vector = (\vector_1, \dots, \vector_\vectorLength) \in \R^\vectorLength$.
\Ensure $\encOffsetVec$ encryption of a vector in $\R^\vectorLength$ representing the tie-correction offset vector of $\vector$.
\State $\encEquality \gets \eq(\encVector)$
\State $\text{mask} \gets \delta_{j \ge i}$
\State $\encScalingEquality \gets \sumR(\encEquality \cdot \text{mask})$
\State $\encTieSize \gets \sumR(\encEquality)$
\State $\encOffsetVec \gets \encScalingEquality - 0.5 \cdot \encTieSize - (0.5, \dots, 0.5)$
\State \Return $\encOffsetVec$
\end{algorithmic}
\end{algorithm}

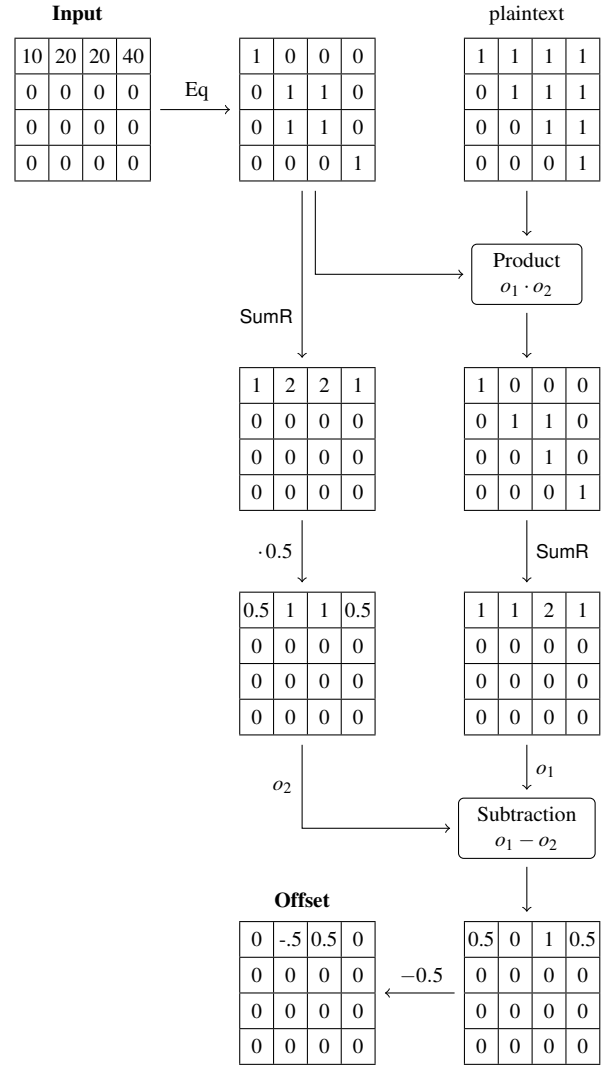
\begin{figure}[!h]
    \centering
    \begin{tikzpicture}[node distance=85pt]

    \tikzstyle{every node}=[font=\footnotesize]
    
    \node (table1) {
    \renewcommand{\tabcolsep}{\tabcolsepValue}
    \renewcommand{\arraystretch}{\arraystretchValue}
    \begin{tabular}{|P{10pt}|P{10pt}|P{10pt}|P{10pt}|}
        \hline
        10 & 20 & 20 & 40 \\ \hline
        0 & 0 & 0 & 0 \\ \hline
        0 & 0 & 0 & 0 \\ \hline
        0 & 0 & 0 & 0 \\ \hline
    \end{tabular}
    };

    \node[right of=table1] (table2) {
    \renewcommand{\tabcolsep}{\tabcolsepValue}
    \renewcommand{\arraystretch}{\arraystretchValue}
    \begin{tabular}{|P{10pt}|P{10pt}|P{10pt}|P{10pt}|}
        \hline
        1 & 0 & 0 & 0 \\ \hline
        0 & 1 & 1 & 0 \\ \hline
        0 & 1 & 1 & 0 \\ \hline
        0 & 0 & 0 & 1 \\ \hline
    \end{tabular}
    };

    \node[right of=table2] (table3) {
    \renewcommand{\tabcolsep}{\tabcolsepValue}
    \renewcommand{\arraystretch}{\arraystretchValue}
    \begin{tabular}{|P{10pt}|P{10pt}|P{10pt}|P{10pt}|}
        \hline
        1 & 1 & 1 & 1 \\ \hline
        0 & 1 & 1 & 1 \\ \hline
        0 & 0 & 1 & 1 \\ \hline
        0 & 0 & 0 & 1 \\ \hline
    \end{tabular}
    };

    \node[draw, rounded corners=2pt, below=20pt of table3, text width=35pt, align=center] (mult) {
    Product $o_1 \cdot o_2$
    };

    \node[below=20pt of mult] (table5) {
    \renewcommand{\tabcolsep}{\tabcolsepValue}
    \renewcommand{\arraystretch}{\arraystretchValue}
    \begin{tabular}{|P{10pt}|P{10pt}|P{10pt}|P{10pt}|}
        \hline
        1 & 0 & 0 & 0 \\ \hline
        0 & 1 & 1 & 0 \\ \hline
        0 & 0 & 1 & 0 \\ \hline
        0 & 0 & 0 & 1 \\ \hline
    \end{tabular}
    };

    \node[left of=table5] (table6) {
    \renewcommand{\tabcolsep}{\tabcolsepValue}
    \renewcommand{\arraystretch}{\arraystretchValue}
    \begin{tabular}{|P{10pt}|P{10pt}|P{10pt}|P{10pt}|}
        \hline
        1 & 2 & 2 & 1 \\ \hline
        0 & 0 & 0 & 0 \\ \hline
        0 & 0 & 0 & 0 \\ \hline
        0 & 0 & 0 & 0 \\ \hline
    \end{tabular}
    };

    \node[below of=table6] (table7) {
    \renewcommand{\tabcolsep}{\tabcolsepValue}
    \renewcommand{\arraystretch}{\arraystretchValue}
    \begin{tabular}{|P{10pt}|P{10pt}|P{10pt}|P{10pt}|}
        \hline
        0.5 & 1 & 1 & 0.5 \\ \hline
        0 & 0 & 0 & 0 \\ \hline
        0 & 0 & 0 & 0 \\ \hline
        0 & 0 & 0 & 0 \\ \hline
    \end{tabular}
    };

    \node[below of=table5] (table8) {
    \renewcommand{\tabcolsep}{\tabcolsepValue}
    \renewcommand{\arraystretch}{\arraystretchValue}
    \begin{tabular}{|P{10pt}|P{10pt}|P{10pt}|P{10pt}|}
        \hline
        1 & 1 & 2 & 1 \\ \hline
        0 & 0 & 0 & 0 \\ \hline
        0 & 0 & 0 & 0 \\ \hline
        0 & 0 & 0 & 0 \\ \hline
    \end{tabular}
    };

    \node[draw, rounded corners=2pt, below=20pt of table8, text width=45pt, align=center] (minus) {
    Subtraction $o_1 - o_2$
    };

    \node[below=20pt of minus] (table9) {
    \renewcommand{\tabcolsep}{\tabcolsepValue}
    \renewcommand{\arraystretch}{\arraystretchValue}
    \begin{tabular}{|P{10pt}|P{10pt}|P{10pt}|P{10pt}|}
        \hline
        0.5 & 0 & 1 & 0.5 \\ \hline
        0 & 0 & 0 & 0 \\ \hline
        0 & 0 & 0 & 0 \\ \hline
        0 & 0 & 0 & 0 \\ \hline
    \end{tabular}
    };

    \node[left of=table9] (table10) {
    \renewcommand{\tabcolsep}{\tabcolsepValue}
    \renewcommand{\arraystretch}{\arraystretchValue}
    \begin{tabular}{|P{10pt}|P{10pt}|P{10pt}|P{10pt}|}
        \hline
        0 & -.5 & 0.5 & 0 \\ \hline
        0 & 0 & 0 & 0 \\ \hline
        0 & 0 & 0 & 0 \\ \hline
        0 & 0 & 0 & 0 \\ \hline
    \end{tabular}
    };

    \node[above=-2pt of table1] {\textbf{Input}};
    \node[above=-2pt of table10] {\textbf{Offset}};
    \node[above=-2pt of table3, text width=35pt, align=center] {plaintext};
    \draw[->, shorten >=-4pt] (table1) -- (table2) node[pos=0.62, above] {$\eq$};
    \begin{scope}[transform canvas={xshift=5pt}]
        \draw[->, shorten >=8pt] (table2) |- (mult) node[midway, right] {};
    \end{scope}
    \draw[->, shorten >=3pt] (table3) -- (mult) node[midway, right] {};
    \draw[->] (table2) -- (table6) node[near end, left] {\sumR};
    \draw[->, shorten <=3pt] (mult) -- (table5) node[midway, right] {};
    \draw[->] (table6) -- (table7) node[midway, left] {$\cdot \, 0.5$};
    \draw[->] (table5) -- (table8) node[midway, right] {\sumR};
    \draw[->, shorten >=3pt] (table7) |- (minus) node[near start, left] {$o_2$};
    \draw[->, shorten >=3pt] (table8) -- (minus) node[midway, right] {$o_1$};
    \draw[->, shorten <=3pt] (minus) -- (table9) node[midway, right] {};
    \draw[->, shorten <=-4pt] (table9) -- (table10) node[pos=0.38, above] {$-0.5$};

    \end{tikzpicture}
    
    \caption{Schematic example of computing the tie-correction offset for a 4-element vector.}
    \label{fig:schematicExampleTieCorrectionOffset}
\end{figure}

\paragraph{Correctness Proof}
Given a vector $\vector \in \R^\vectorLength$, let $\encVector$ be its encryption.
Let $\encRanking$ and $\encOffsetVec$ be the output of \Cref{alg:ranking} and \Cref{alg:tie-correction-offset} on input $\encVector$, respectively.
And let $\ranking$ and $\offsetVec$ be the corresponding decryption.
Assuming the correctness of the building blocks and \Cref{alg:ranking}, and the ideal functionality of the algorithms, we prove that (1) $k := \ranking + \offsetVec$ is a permutation of $(1, \dots, \vectorLength)$, and (2) $\vector_j$ is the $k_j$-th order statistic of $\vector$, for all $j \in \{1, \dots, \vectorLength\}$.
Let $\equality$, $\scalingEquality$, and $\tieSize$ be the decryption of the intermediate computations $\encEquality$, $\encScalingEquality$, and $\encTieSize$ in \Cref{alg:tie-correction-offset}, respectively.
Let $j \in \{1, \dots, \vectorLength\}$, and let us define the following subsets of $\{1, \dots, \vectorLength\}$:
\begin{align*}
    \partitionLower_j &:= \{ i \in \{1,\dots,\vectorLength\} : v_i < v_j\} \\
    \partitionEqual_j &:= \{ i \in \{1,\dots,\vectorLength\} : v_i = v_j\} \\
    \partitionEqualScaled_j &:= \{ i \in \{1,\dots,j\} : v_i = v_j\} \enspace .
\end{align*}
By the correctness of $\sumR$, we have that
\begin{align*}
    \scalingEquality_j &= \sum_{i = 1}^{\vectorLength}{\equality_{ij} \delta_{j \ge i}} = |\partitionEqualScaled_j| \\
    \tieSize_j &= \sum_{i = 1}^{\vectorLength}{\equality_{ij}} = |\partitionEqual_j|
\end{align*}
and thus
\begin{align*}
    \offsetVec_j &= \scalingEquality_j - 0.5 \cdot \tieSize_j - 0.5 \\
    &= |\partitionEqualScaled_j| - 0.5 \cdot |\partitionEqual_j| - 0.5 \enspace .
\end{align*}
On the other hand, by the correctness of \Cref{alg:ranking}, we know that
$$\ranking_j = |\partitionLower_j| + 0.5 \cdot \left(|\partitionEqual_j| + 1\right) \enspace .$$
Combining these two identities we get
\begin{align*}
    k_j :=& \,\ranking_j + \offsetVec_j \\
    =& \,|\partitionEqualScaled_j| - 0.5 \cdot |\partitionEqual_j| - 0.5 + |\partitionLower_j| + 0.5 \cdot \left(|\partitionEqual_j| + 1\right) \\
    =& \,|\partitionEqualScaled_j| + |\partitionLower_j| \enspace .
\end{align*}
Let $w$ be the sorted array, then we note that $w_i = v_j$ for all $i \in \{ |\partitionLower_j| + 1, \dots, |\partitionLower_j| + |\partitionEqual_j|\}$.
Since $|\partitionEqualScaled_j| \ge 1$ (as it contains at least the trivial identity), and $|\partitionEqualScaled_j| \le |\partitionEqual_j|$ (as $\partitionEqualScaled_j \subseteq \partitionEqual_j$), we have that $|\partitionLower_j| + 1 \le k \le |\partitionLower_j| + |\partitionEqual_j|$.
Hence, $w_{k_j} = v_j$, proving point (2).

Now, we prove that $k$ is a permutation of $(1, \dots, \vectorLength)$. Let $j \in \{1, \dots, \vectorLength\}$, then
\begin{itemize}
    \item $k_j \in \N$: this is trivial, since both $|\partitionEqualScaled_j|, |\partitionLower_j| \in \N$;
    \item $k_j \ge 1$: since $v_j = v_j$, we have that $k_j \ge |\partitionEqualScaled_j| \ge 1$;
    \item $k_j \le \vectorLength$: this is true since $\partitionEqualScaled_j$ and $\partitionLower_j$ are non-overlapping subsets of $\{1, \dots, \vectorLength\}$;
    \item for all $j' \ne j$, $k_{j'} \ne k_j$: we consider three cases
    \begin{enumerate}
        \item if $v_j = v_{j'}$, then $\partitionLower_j = \partitionLower_{j'}$; without loss of generality, we assume $j' > j$, thus $|\partitionEqualScaled_{j'}| > |\partitionEqualScaled_j|$, hence $k_{j'} > k_j$;
        \item if $v_j < v_{j'}$, then $\partitionLower_j \cup \partitionEqualScaled_j \subseteq \partitionLower_j \cup \partitionEqual_j \subseteq \partitionLower_{j'}$, thus $k_j = |\partitionLower_j| + |\partitionEqualScaled_j| < |\partitionLower_{j'}| + 1 \le k_{j'}$;
        \item if $v_j > v_{j'}$, the proof is symmetric to the previous case.
    \end{enumerate}
\end{itemize}
This proves point (1).
\qed


\section{Multiple Ciphertexts Encoding}
\label{sec:multiple_ciphertexts}

\begin{figure}
    \centering
    \begin{tabular}{ccccccccc}

    \renewcommand{\arraystretch}{1.03}
    \renewcommand{\tabcolsep}{\tabcolsepValue}
    \begin{tabular}{|P{9pt}|P{11pt}|P{9pt}|P{9pt}|P{9pt}|P{9pt}||P{9pt}|P{9pt}|P{9pt}|P{9pt}||P{9pt}|P{9pt}|P{9pt}|P{9pt}||P{9pt}|P{9pt}|P{9pt}|P{9pt}|}
        \cline{3-18}
        \multicolumn{2}{P{20pt}|}{} & $\vector_1$ & $\vector_2$ & $\vector_3$ & $\vector_4$ & %
        $\vector_5$ & $\vector_6$ & $\vector_7$ & $\vector_8$ & %
        \multicolumn{4}{P{44pt}||}{$\dots$} & %
        \multicolumn{3}{P{33pt}|}{$\dots$} & $\!\vector_\vectorLength$ \\ \cline{3-18}
        \multicolumn{2}{P{20pt}|}{} & \multicolumn{4}{P{44pt}||}{$\encBlock_1$} & %
        \multicolumn{4}{P{44pt}||}{$\encBlock_2$} & %
        \multicolumn{4}{P{44pt}||}{$\dots$} & %
        \multicolumn{4}{P{44pt}|}{$\encBlock_\numBlocks$} \\ \hline
        $\vector_1$ & \multirow{4}{*}{$\encBlock_1$} & %
        \multicolumn{4}{P{44pt}||}{\cellcolor{blue!10}} & 
        \multicolumn{4}{P{44pt}||}{\cellcolor{blue!10}} & 
        \multicolumn{4}{P{44pt}||}{\cellcolor{blue!10}} &
        \multicolumn{4}{P{44pt}|}{\cellcolor{blue!10}} \\ \cline{1-1}
        $\vector_2$ & & %
        \multicolumn{4}{P{44pt}||}{\cellcolor{blue!10}} & 
        \multicolumn{4}{P{44pt}||}{\cellcolor{blue!10}} & 
        \multicolumn{4}{P{44pt}||}{\cellcolor{blue!10}} &
        \multicolumn{4}{P{44pt}|}{\cellcolor{blue!10}} \\ \cline{1-1}
        $\vector_3$ & & %
        \multicolumn{4}{P{44pt}||}{\cellcolor{blue!10}} & 
        \multicolumn{4}{P{44pt}||}{\cellcolor{blue!10}} & 
        \multicolumn{4}{P{44pt}||}{\cellcolor{blue!10}} &
        \multicolumn{4}{P{44pt}|}{\cellcolor{blue!10}} \\ \cline{1-1}
        $\vector_4$ & & %
        \multicolumn{4}{P{44pt}||}{\cellcolor{blue!10}\multirow{-4}{*}{$\encBlock_1 > \encBlock_1$}} & 
        \multicolumn{4}{P{44pt}||}{\cellcolor{blue!10}\multirow{-4}{*}{$\encBlock_1 > \encBlock_2$}} & 
        \multicolumn{4}{P{44pt}||}{\cellcolor{blue!10}\multirow{-4}{*}{$\dots$}} &
        \multicolumn{4}{P{44pt}|}{\cellcolor{blue!10}\multirow{-4}{*}{$\encBlock_1 > \encBlock_\numBlocks$}} \\ \hline\hline
        $\vector_5$ & \multirow{4}{*}{$\encBlock_2$} & %
        \multicolumn{4}{P{44pt}||}{} & 
        \multicolumn{4}{P{44pt}||}{\cellcolor{blue!10}} & 
        \multicolumn{4}{P{44pt}||}{\cellcolor{blue!10}} &
        \multicolumn{4}{P{44pt}|}{\cellcolor{blue!10}} \\ \cline{1-1}
        $\vector_6$ & & %
        \multicolumn{4}{P{44pt}||}{} & 
        \multicolumn{4}{P{44pt}||}{\cellcolor{blue!10}} & 
        \multicolumn{4}{P{44pt}||}{\cellcolor{blue!10}} &
        \multicolumn{4}{P{44pt}|}{\cellcolor{blue!10}} \\ \cline{1-1}
        $\vector_7$ & & %
        \multicolumn{4}{P{44pt}||}{} & 
        \multicolumn{4}{P{44pt}||}{\cellcolor{blue!10}} & 
        \multicolumn{4}{P{44pt}||}{\cellcolor{blue!10}} &
        \multicolumn{4}{P{44pt}|}{\cellcolor{blue!10}} \\ \cline{1-1}
        $\vector_8$ & & %
        \multicolumn{4}{P{44pt}||}{\multirow{-4}{*}{$\encBlock_2 > \encBlock_1$}} & 
        \multicolumn{4}{P{44pt}||}{\cellcolor{blue!10}\multirow{-4}{*}{$\encBlock_2 > \encBlock_2$}} & 
        \multicolumn{4}{P{44pt}||}{\cellcolor{blue!10}\multirow{-4}{*}{$\dots$}} &
        \multicolumn{4}{P{44pt}|}{\cellcolor{blue!10}\multirow{-4}{*}{$\encBlock_2 > \encBlock_\numBlocks$}} \\ \hline\hline
        \multirow{4}{*}{$\vdots$} & \multirow{4}{*}{$\vdots$} & %
        \multicolumn{4}{P{44pt}||}{} & 
        \multicolumn{4}{P{44pt}||}{} & 
        \multicolumn{4}{P{44pt}||}{\cellcolor{blue!10}} &
        \multicolumn{4}{P{44pt}|}{\cellcolor{blue!10}} \\
         & & %
        \multicolumn{4}{P{44pt}||}{} & 
        \multicolumn{4}{P{44pt}||}{} & 
        \multicolumn{4}{P{44pt}||}{\cellcolor{blue!10}} &
        \multicolumn{4}{P{44pt}|}{\cellcolor{blue!10}} \\
         & & %
        \multicolumn{4}{P{44pt}||}{} & 
        \multicolumn{4}{P{44pt}||}{} & 
        \multicolumn{4}{P{44pt}||}{\cellcolor{blue!10}} &
        \multicolumn{4}{P{44pt}|}{\cellcolor{blue!10}} \\
         & & %
        \multicolumn{4}{P{44pt}||}{\multirow{-4}{*}{$\vdots$}} & 
        \multicolumn{4}{P{44pt}||}{\multirow{-4}{*}{$\vdots$}} & 
        \multicolumn{4}{P{44pt}||}{\cellcolor{blue!10}\multirow{-4}{*}{$\ddots$}} &
        \multicolumn{4}{P{44pt}|}{\cellcolor{blue!10}\multirow{-4}{*}{$\vdots$}} \\ \hline\hline
        \multirow{3}{*}{$\vdots$} & \multirow{4}{*}{$\encBlock_\numBlocks$} & %
        \multicolumn{4}{P{44pt}||}{} & 
        \multicolumn{4}{P{44pt}||}{} & 
        \multicolumn{4}{P{44pt}||}{} &
        \multicolumn{4}{P{44pt}|}{\cellcolor{blue!10}} \\
         & & %
        \multicolumn{4}{P{44pt}||}{} & 
        \multicolumn{4}{P{44pt}||}{} & 
        \multicolumn{4}{P{44pt}||}{} &
        \multicolumn{4}{P{44pt}|}{\cellcolor{blue!10}} \\
         & & %
        \multicolumn{4}{P{44pt}||}{} & 
        \multicolumn{4}{P{44pt}||}{} & 
        \multicolumn{4}{P{44pt}||}{} &
        \multicolumn{4}{P{44pt}|}{\cellcolor{blue!10}} \\ \cline{1-1}
        $\!\vector_\vectorLength$ & & %
        \multicolumn{4}{P{44pt}||}{\multirow{-4}{*}{$\encBlock_\numBlocks > \encBlock_1$}} & 
        \multicolumn{4}{P{44pt}||}{\multirow{-4}{*}{$\encBlock_\numBlocks > \encBlock_2$}} & 
        \multicolumn{4}{P{44pt}||}{\multirow{-4}{*}{$\dots$}} &
        \multicolumn{4}{P{44pt}|}{\cellcolor{blue!10}\multirow{-4}{*}{$\encBlock_\numBlocks > \encBlock_\numBlocks$}} \\ \hline
    \end{tabular}
    &
    \end{tabular}
    \caption{Comparison in multi-ciphertext mode. The vector $\vector$ is split into blocks of size $\blockSize = 4$. The upper triangle contains information about the comparison between all pairs $\vector_i > \vector_j$.}
    \label{fig:multiCiphertextComparison}
\end{figure}

When a vector is too long and does not fit in a ciphertext we can split it into multiple blocks.
This happens when $\vectorLength^2 > n/2$, where $n$ is the ring dimension and $\vectorLength$ is the vector length.
In this case, we divide the vector into $\numBlocks$ blocks $\encBlock_1, \dots, \encBlock_\numBlocks$ of size $\blockSize = 2^{\floor{\log{\sqrt{n/2}}}}$, which can be done under encryption by suitable rotations and masking.
When it comes to comparisons, we also have to consider the comparisons between different blocks, that is, computing
$$\encComparison_{i,j} = (\encBlock_i > \encBlock_j) := \cmp(\replR(\encBlock_i), \replC(\transR(\encBlock_j))$$
for all $i,j \in \{1, \dots, \numBlocks\}$.
The results can then be aggregated row-wise and block-wise to compute the ranking
$$\encRanking_i = \sum_{j = 1}^{\numBlocks}{\sumR(\encComparison_{i,j})} + 0.5$$
for $i \in \{1, \dots, \numBlocks\}$. Note that we can also compute the block-sum first, as $\sum_{j = 1}^{\numBlocks}{\sumR(\encComparison_{i,j})} = \sumR(\sum_{j = 1}^{\numBlocks}{\encComparison_{i,j}})$, allowing for evaluating \sumR \, only once per block.
The split output $\encRanking_1,\dots,\encRanking_\numBlocks$ can then be merged back into a single ciphertext if needed and if it fits.

The total number of comparisons is $\numBlocks^2$, but they can all be computed in parallel, making this approach suitable for a multi-threaded environment.
To reduce the computational burden, we notice that not all the comparisons $\encComparison_{i,j}$ are actually needed, as the information in $\encComparison_{i,j}$ is already contained in $\encComparison_{j,i}$ for all $i, j$ (see \Cref{fig:multiCiphertextComparison}).
In particular, we have that
\begin{equation}
\label{eq:comparisonComplement}
    \encComparison_{i,j} = (1 - \encComparison_{j,i})^\top .
\end{equation}

\paragraph{Proof of Equation~\ref{eq:comparisonComplement}}
For the sake of notation, let $A := \encComparison_{i,j}$ and $B := \encComparison_{j,i}$.
Then $A_{m,n} = \cmp(\encBlock_{i;n}, \encBlock_{j;m})$, where $\encBlock_{x;y}$ denotes the $y$-th element of block $x$.
On the other hand, $B_{n,m} = \cmp(\encBlock_{j;m}, \encBlock_{i;n}) = 1 - \cmp(\encBlock_{i;n}, \encBlock_{j;m})$.
\qed

\vspace{4pt}
Hence, we compute $\encComparison_{i,j}$ only for $j \ge i$ and use Equation~\ref{eq:comparisonComplement} for $j < i$.
To avoid the transposition of $1 - \encComparison_{j,i}$ as a whole matrix, which is expensive, we operate on it column-wise, and only transpose it in the end, after summing it up to a vector:
$$\encRanking_i =
\transC\Big(\sumC\Big(\sum_{j = 1}^{i - 1}{(1 - \encComparison_{j,i})}\Big)\Big) +
\sumR\Big(\sum_{j = i}^{\numBlocks}{\encComparison_{i,j}}\Big) .$$
This optimization makes us save $\numBlocks (\numBlocks - 1) / 2$ comparisons.
\Cref{alg:multiCtxt_ranking} describes the full pseudocode for multi-ciphertext ranking.
The correctness can be proven similarly as for \Cref{alg:ranking}, by exploiting Equation~\ref{eq:comparisonComplement}.
The algorithm in case of ranking with tie-correction is similar.
We omit its description in the multi-ciphertext pseudocode for the sake of clarity.

\begin{algorithm}[t]
\caption{Multi-Ciphertext Ranking}
\label{alg:multiCtxt_ranking}
\begin{algorithmic}[1]
\Require $\encBlock_1, \dots, \encBlock_\numBlocks$ multi-ciphertext encryption of $\vector = (\vector_1, \dots, \vector_\vectorLength) \in \R^\vectorLength$, approximation degree $\approximationDegree \in \N$.
\Ensure $\encRanking_1, \dots, \encRanking_\numBlocks$ multi-ciphertext encryption of a vector in $\R^\vectorLength$ representing the (fractional) ranking of $\vector$.
\ParFor{$i = 1, \dots, \numBlocks$}
    \State $\encBlock_{R;i} \gets \replR(\encBlock_i)$
    \State $\encBlock_{C;i} \gets \replC(\transR(\encBlock_i))$
\EndParFor
\ParFor{$i = 1, \dots, \numBlocks$}
    \ParFor{$j = i, \dots, \numBlocks$}
        \State $\encComparison_{i,j} \gets \cmp(\encBlock_{R;i}, \encBlock_{C;j}; \approximationDegree)$
    \EndParFor
\EndParFor
\ParFor{$i = 1, \dots, \numBlocks$}
    \State $\encRanking_i \gets \transC(\sumC(\sum_{j = 1}^{i - 1}{(1 - \encComparison_{j,i})})) + $ $\hphantom{\transC}\sumR(\sum_{j = i}^{\numBlocks}{\encComparison_{i,j}}) + (0.5, \dots, 0.5)$
\EndParFor
\State \Return $\encRanking_1, \dots, \encRanking_\numBlocks$
\end{algorithmic}
\end{algorithm}

\begin{algorithm}
\caption{Multi-Ciphertext Sorting}
\label{alg:multiCtxt_sorting}
\begin{algorithmic}[1]
\Require $\encBlock_1, \dots, \encBlock_\numBlocks$ multi-ciphertext encryption of $\vector = (\vector_1, \dots, \vector_\vectorLength) \in \R^\vectorLength$ with distinct elements, approximation degrees $\approximationDegree_C, \approximationDegree_I \in \N$.
\Ensure $\encSortedVec_1, \dots, \encSortedVec_\numBlocks$ multi-ciphertext encryption of the sorted form of $\vector$.
\State $\encRanking_1, \dots, \encRanking_\numBlocks \gets \rank(\encBlock_1, \dots, \encBlock_\numBlocks; \approximationDegree_C)$
\ParFor{$i = 1, \dots, \numBlocks$}
    \State $\encRanking_{R;i} \gets \replR(\encRanking_i)$
\EndParFor
\ParFor{$i = 1, \dots, \numBlocks$}
    \State $\encBlock_{R;i} \gets \replR(\encBlock_i)$
    \State $\encSortedVec_i \gets 0$
    \ParFor{$j = 1, \dots, \numBlocks$}
        \State $\encSortingMask_{i,j} \gets \ind_0(\encRanking_{R;j} - ((\blockSize(i-1)+1)^\vectorLength \parallel \cdots \parallel$
        $\hphantom{\encSortingMask_{i,j} \gets \ind_0(\encRanking_{R;j}}(\blockSize i)^\vectorLength); \approximationDegree_I)$
        \State $\encSortedVec_i \gets \encSortedVec_i + \encSortingMask_{i,j} \cdot \encVector_{R;j}$
    \EndParFor
    \State $\encSortedVec_i \gets \transC(\sumC(\encSortedVec_i))$
\EndParFor
\State \Return $\encSortedVec_1, \dots, \encSortedVec_\numBlocks$
\end{algorithmic}
\end{algorithm}

\begin{figure*}[!t]
    \newcommand{\figurewidth}{230pt}
    \newcommand{\figureheight}{195pt}
    \newcommand{\xlabelshift}{5pt}
    \newcommand{\ylabelshift}{-10pt}
    \newcommand{\stlinestyle}{densely dashed}
    \centering
    \begin{subfigure}[t]{\columnwidth}
        \centering
        \begin{tikzpicture}
            \begin{axis}[
                width=\figurewidth,
                height=\figureheight,
                xlabel=vector length,
                ylabel=runtime (s),
                grid=both,
                xmode=log,
                ymode=log,
                log basis x=2,
                log basis y=2,
                log ticks with fixed point,
                xtick={8,16,32,64,128,256,512,1024,2048,4096,8192,16384},
                xticklabels={8,16,32,64,128,256,512,1024,2048,4096,8192,16384},
                xticklabel style={rotate=45, anchor=north east},
                ytick={0.5, 2, 8, 32, 128, 512, 2048, 8192, 32768, 131072},
                yticklabels={0.5, 2, 8, 32, 128, 512, 2048, 8192, 32768, 131072},
                enlarge x limits=0.03,
                enlarge y limits=0.05,
                label style={font=\small},
                tick label style={font=\small},
                legend style={
                    at={(0.5,1.1)},
                    anchor=south,
                    font=\footnotesize,
                    yshift=\ylabelshift,
                    draw=none
                },
                legend columns=2,
                legend cell align={left},
                xlabel style={yshift=\xlabelshift},
                ylabel style={yshift=\ylabelshift}
                ]
        
            \addplot[thick,color=supergreen!80,mark=*,mark size=1.3pt]
                plot coordinates {
                    (8, 0.56)
                    (16, 0.79)
                    (32, 3.64)
                    (64, 3.69)
                    (128, 5.76)
                    (256, 12.77)
                    (512, 31.36)
                    (1024, 32.09)
                    (2048, 84.12)
                    (4096, 253.34)
                    (8192, 871.3)
                    (16384, 3260.83)
                };
            \addplot[thick,color=violet!70,mark=triangle*,mark size=1.3pt]
                plot coordinates {
                    (8, 1.66)
                    (16, 3.64)
                    (32, 4.7)
                    (64, 5.19)
                    (128, 13.18)
                    (256, 25.97)
                    (512, 38.38)
                    (1024, 40.81)
                    (2048, 106.19)
                    (4096, 316.15)
                    (8192, 1088.76)
                    (16384, 4076.07)
                };
            \addplot[\stlinestyle,very thick,color=supergreen!80,mark=*,mark size=1.3pt]
                plot coordinates {
                    (8, 1.21)
                    (16, 1.77)
                    (32, 8.36)
                    (64, 8.59)
                    (128, 13.75)
                    (256, 43.04)
                    (512, 215.81)
                    (1024, 685.42)
                    (2048, 2367.47)
                    (4096, 8635.78)
                    (8192, 33237.43)
                    (16384, 129946.68)
                };
            \addplot[very thick,\stlinestyle,color=violet!70,mark=triangle*,mark size=1.3pt]
                plot coordinates {
                    (8, 4.31)
                    (16, 8.69)
                    (32, 11.98)
                    (64, 12.50)
                    (128, 32.77)
                    (256, 105.22)
                    (512, 279.10)
                    (1024, 866.99)
                    (2048, 2965.93)
                    (4096, 10826.89)
                    (8192, 41222.92)
                    (16384, 160706.46)
                };

            \addlegendentry{our solution (basic) MT}
            \addlegendentry{our solution (tie-corr.) MT}
            \addlegendentry{our solution (basic) ST}
            \addlegendentry{our solution (tie-corr.) ST}
        
            \draw[very thin, dashed] (axis cs:181.019, 0.298741641) -- (axis cs:181.019, 301248.9904);
            \node[font=\footnotesize,text width=30pt,align=right,anchor=north east] at (170,160706.46) {single-ciphertext};
            \node[font=\footnotesize,text width=30pt,align=left,anchor=north west] at (192,160706.46) {multi-ciphertext};
            \draw[->] (156,160706.46) -- (104,160706.46);
            \draw[->] (208,160706.46) -- (315,160706.46);
        
            \draw[very thin, dashed] (axis cs:1448.155, 0.298741641) -- (axis cs:1448.155, 301248.9904) node[pos=0.14,left,rotate=90,anchor=north,font=\footnotesize] {CPU limit};
            \end{axis}
        \end{tikzpicture}
        \caption{Ranking}
        \vspace{18pt}
        \label{fig:runtime:rank}
    \end{subfigure}
    \begin{subfigure}[t]{\columnwidth}
        \centering
        \begin{tikzpicture}
            \begin{axis}[
                width=\figurewidth,
                height=\figureheight,
                xlabel=vector length,
                ylabel=runtime (s),
                grid=both,
                xmode=log,
                ymode=log,
                log basis x=2,
                log basis y=2,
                log ticks with fixed point,
                xtick={8,16,32,64,128,256,512,1024,2048,4096,8192,16384},
                xticklabels={8,16,32,64,128,256,512,1024,2048,4096,8192,16384},
                xticklabel style={rotate=45, anchor=north east},
                ytick={0.5, 2, 8, 32, 128, 512, 2048, 8192, 32768, 131072, 524288, 2097152},
                yticklabels={0.5, 2, 8, 32, 128, 512, 2048, 8192, 32768, 131072, 524288, 2097152},
                enlarge x limits=0.03,
                enlarge y limits=0.05,
                label style={font=\small},
                tick label style={font=\small},
                legend pos=south east,
                legend cell align={left},
                legend style={
                    at={(0.5,1.1)},
                    anchor=south,
                    font=\footnotesize,
                    yshift=\ylabelshift,
                    draw=none
                },
                legend columns=3,
                legend cell align={left},
                xlabel style={yshift=\xlabelshift},
                ylabel style={yshift=\ylabelshift}
                ]
        
            \addplot[thick,color=supergreen!80,mark=*,mark size=1.3pt]
                plot coordinates {
                    (8, 6.14)
                    (16, 6.80)
                    (32, 7.18)
                    (64, 10.36)
                    (128, 12.83)
                    (256, 24.99)
                    (512, 83.24)
                    (1024, 179.56)
                    (2048, 255.95)
                    (4096, 806.09)
                    (8192, 2935.18)
                    (16384, 12537.34)
                };
            \addplot[thick,color=blue!70,mark=triangle*,mark size=1.3pt]
                plot coordinates {
                    (8, 117.65)
                    (16, 145.18)
                    (32, 175.97)
                    (64, 263.94)
                    (128, 302.88)
                    (256, 336.41)
                    (512, 473.59)
                    (1024, 518.75)
                    (2048, 570.54)
                    (4096, 638.65)
                    (8192, 698.57)
                    (16384, 745.76)
                };
            \addplot[thick,color=red!70,mark=square*,mark size=1.3pt]
                plot coordinates {
                    (8, 313.73)
                    (16, 580.72)
                    (32, 1126.21)
                    (64, 2815.36)
                    (128, 5538.38)
                    (256, 10765.12)
                    (512, 26942.01)
                    (1024, 53120.00)
                    (2048, 106224.17)
                    (4096, 217992.53)
                    (8192, 440206.57)
                    (16384, 872752.27)
                };
            \addplot[\stlinestyle,very thick,color=supergreen!80,mark=*,mark size=1.3pt]
                plot coordinates {
                    (8, 15.65)
                    (16, 16.87)
                    (32, 18.01)
                    (64, 30.09)
                    (128, 31.95)
                    (256, 112.61)
                    (512, 648.55)
                    (1024, 1994.30)
                    (2048, 7032.72)
                    (4096, 29354.27)
                    (8192, 113899.77)
                    (16384, 498145.97)
                };
            \addplot[\stlinestyle,very thick,color=blue!70,mark=triangle*,mark size=1.3pt]
                plot coordinates {
                    (8, 457.75)
                    (16, 571.46)
                    (32, 688.56)
                    (64, 1026.96)
                    (128, 1181.35)
                    (256, 1311.53)
                    (512, 2114.02)
                    (1024, 2340.60)
                    (2048, 2605.94)
                    (4096, 2865.70)
                    (8192, 3153.06)
                    (16384, 3472.11)
                };
            \addplot[\stlinestyle,very thick,color=red!70,mark=square*,mark size=1.3pt]
                plot coordinates {
                    (8, 1220.67)
                    (16, 2285.84)
                    (32, 4406.78)
                    (64, 10954.24)
                    (128, 21601.83)
                    (256, 41968.96)
                    (512, 120264.25)
                    (1024, 239677.44)
                    (2048, 485178.65)
                    (4096, 978159.15)
                    (8192, 1986914.51)
                    (16384, 4063363.59)
                };
            \addlegendentry{our solution MT}
            \addlegendentry{NEXUS MT}
            \addlegendentry{Phoenix MT}
            \addlegendentry{our solution ST}
            \addlegendentry{NEXUS ST}
            \addlegendentry{Phoenix ST}
            
            \draw[very thin, dashed] (axis cs:181.019, 3.141467025) -- (axis cs:181.019, 7941847.628);
            \node[font=\footnotesize,text width=30pt,align=right,anchor=north east] at (170,4063363.59) {single-ciphertext};
            \node[font=\footnotesize,text width=30pt,align=left,anchor=north west] at (192,4063363.59) {multi-ciphertext};
            \draw[->] (156,4063363.59) -- (104,4063363.59);
            \draw[->] (208,4063363.59) -- (315,4063363.59);
        
            \draw[very thin, dashed] (axis cs:2896.30937574, 3.141467025) -- (axis cs:2896.30937574, 7941847.628) node[pos=0.14,left,rotate=90,anchor=north,font=\footnotesize] {CPU limit};
        
            \end{axis}
        \end{tikzpicture}
        \caption{Minimum}
        \vspace{18pt}
        \label{fig:runtime:min}
    \end{subfigure}
    \begin{subfigure}[t]{\columnwidth}
        \centering
        \begin{tikzpicture}
            \begin{axis}[
                width=\figurewidth,
                height=\figureheight,
                xlabel=vector length,
                ylabel=runtime (s),
                grid=both,
                xmode=log,
                ymode=log,
                log basis x=2,
                log basis y=2,
                log ticks with fixed point,
                xtick={8,16,32,64,128,256,512,1024,2048,4096,8192,16384},
                xticklabels={8,16,32,64,128,256,512,1024,2048,4096,8192,16384},
                xticklabel style={rotate=45, anchor=north east},
                ytick={0.5, 2, 8, 32, 128, 512, 2048, 8192, 32768, 131072, 524288, 2097152},
                yticklabels={0.5, 2, 8, 32, 128, 512, 2048, 8192, 32768, 131072, 524288, 2097152},
                ymax=600000,
                enlarge x limits=0.03,
                enlarge y limits=0.05,
                label style={font=\small},
                tick label style={font=\small},
                legend pos=south east,
                legend cell align={left},
                legend style={
                    at={(0.5,1.1)},
                    anchor=south,
                    font=\footnotesize,
                    yshift=\ylabelshift,
                    draw=none
                },
                legend columns=2,
                legend cell align={left},
                xlabel style={yshift=\xlabelshift},
                ylabel style={yshift=\ylabelshift}
                ]
        
            \addplot[thick,color=supergreen!80,mark=*,mark size=1.3pt]
                plot coordinates {
                    (8, 27.90)
                    (16, 37.44)
                    (32, 38.94)
                    (64, 50.87)
                    (128, 54.16)
                    (256, 69.09)
                    (512, 88.00)
                    (1024, 174.96)
                    (2048, 251.09)
                    (4096, 722.09)
                    (8192, 1910.72)
                    (16384, 7450.69)
                };
            \addplot[thick,color=blue!70,mark=triangle*,mark size=1.3pt]
                plot coordinates {
                    (8, 1057.09)
                    (16, 1057.09)
                    (32, 2827.91)
                    (64, 2827.91)
                    (128, 5355.55)
                    (256, 5355.55)
                    (512, 5355.55)
                    (1024, 8688.16)
                    (2048, 8688.16)
                    (4096, 13351.63)
                    (8192, 13351.63)
                    (16384, 20340.25)
                };
            \addplot[\stlinestyle,very thick,color=supergreen!80,mark=*,mark size=1.3pt]
                plot coordinates {
                    (8, 97.21)
                    (16, 133.94)
                    (32, 143.31)
                    (64, 188.31)
                    (128, 200.3)
                    (256, 253.16)
                    (512, 638.85)
                    (1024, 1884.87)
                    (2048, 5491.45)
                    (4096, 20049.91)
                    (8192, 70953.20)
                    (16384, 295252.98)
                };
            \addplot[\stlinestyle,very thick,color=blue!70,mark=triangle*,mark size=1.3pt]
                plot coordinates {
                    (8, 13334.34)
                    (16, 13334.34)
                    (32, 34479.67)
                    (64, 34479.67)
                    (128, 66418.06)
                    (256, 66418.06)
                    (512, 66418.06)
                    (1024, 107741.11)
                    (2048, 107741.11)
                    (4096, 165568.33)
                    (8192, 165568.33)
                    (16384, 252247.27)
                };
            \addlegendentry{our solution MT}
            \addlegendentry{Hong et al. MT}
            \addlegendentry{our solution ST}
            \addlegendentry{Hong et al. ST}
        
            \draw[very thin, dashed] (axis cs:362.038671968, 16.9424749) -- (axis cs:362.038671968, 988049.2725);
            \node[font=\footnotesize,text width=30pt,align=right,anchor=north east] at (340.1435684,600000) {single-ciphertext};
            \node[font=\footnotesize,text width=30pt,align=left,anchor=north west] at (385.3431673,600000) {multi-ciphertext};
            \draw[->] (315.1729698,600000) -- (207.9366135,600000);
            \draw[->] (415.8732269,600000) -- (630.3459396,600000);

            \draw[very thin, dashed] (axis cs:2896.30937574, 16.9424749) -- (axis cs:2896.30937574, 988049.2725) node[pos=0.14,left,rotate=90,anchor=north,font=\footnotesize] {CPU limit};

            \end{axis}
        \end{tikzpicture}
        \caption{Median}
        \vspace{18pt}
        \label{fig:runtime:median}
    \end{subfigure}
    \begin{subfigure}[t]{\columnwidth}
        \centering
        \begin{tikzpicture}
            \begin{axis}[
                width=\figurewidth,
                height=\figureheight,
                xlabel=vector length,
                ylabel=runtime (s),
                grid=both,
                xmode=log,
                ymode=log,
                log basis x=2,
                log basis y=2,
                log ticks with fixed point,
                xtick={8,16,32,64,128,256,512,1024,2048,4096,8192,16384},
                xticklabels={8,16,32,64,128,256,512,1024,2048,4096,8192,16384},
                xticklabel style={rotate=45, anchor=north east},
                ytick={0.5, 2, 8, 32, 128, 512, 2048, 8192, 32768, 131072, 524288, 2097152},
                yticklabels={0.5, 2, 8, 32, 128, 512, 2048, 8192, 32768, 131072, 524288, 2097152},
                enlarge x limits=0.03,
                enlarge y limits=0.05,
                label style={font=\small},
                tick label style={font=\small},
                legend pos=south east,
                legend cell align={left},
                legend style={
                    at={(0.5,1.1)},
                    anchor=south,
                    font=\footnotesize,
                    yshift=\ylabelshift,
                    draw=none
                },
                legend columns=2,
                legend cell align={left},
                xlabel style={yshift=\xlabelshift},
                ylabel style={yshift=\ylabelshift}
                ]
        
            \addplot[thick,color=supergreen!80,mark=*,mark size=1.3pt]
                plot coordinates {
                    (8, 43.75)
                    (16, 46.81)
                    (32, 59.70)
                    (64, 63.91)
                    (128, 78.64)
                    (256, 84.00)
                    (512, 155.14)
                    (1024, 344.52)
                    (2048, 564.87)
                    (4096, 1782.78)
                    (8192, 7267.75)
                    (16384, 28065.08)
                };        
            \addplot[thick,color=blue!70,mark=triangle*,mark size=1.3pt]
                plot coordinates {
                    (8, 946.32)
                    (16, 946.32)
                    (32, 2715.55)
                    (64, 2715.55)
                    (128, 5240.80)
                    (256, 5240.80)
                    (512, 5240.80)
                    (1024, 8571.82)
                    (2048, 8571.82)
                    (4096, 13233.70)
                    (8192, 13233.70)
                    (16384, 20221.53)
                };
            \addplot[\stlinestyle,very thick,color=supergreen!80,mark=*,mark size=1.3pt]
                plot coordinates {
                    (8, 133.45)
                    (16, 176.10)
                    (32, 227.28)
                    (64, 240.47)
                    (128, 301.94)
                    (256, 317.06)
                    (512, 1313.19)
                    (1024, 4241.87)
                    (2048, 14555.46)
                    (4096, 53206.78)
                    (8192, 202512.69)
                    (16384, 789137.56)
                };
            \addplot[\stlinestyle,very thick,color=blue!70,mark=triangle*,mark size=1.3pt]
                plot coordinates {
                    (8, 11944.70)
                    (16, 11944.70)
                    (32, 33075.40)
                    (64, 33075.40)
                    (128, 64991.85)
                    (256, 64991.85)
                    (512, 64991.85)
                    (1024, 106300.27)
                    (2048, 106300.27)
                    (4096, 164112.86)
                    (8192, 164112.86)
                    (16384, 250769.86)
                };
            \addlegendentry{our solution MT}
            \addlegendentry{Hong et al. MT}
            \addlegendentry{our solution ST}
            \addlegendentry{Hong et al. ST}
        
            \draw[very thin, dashed] (axis cs:362.038671968, 26.80213093) -- (axis cs:362.038671968, 1288135.198);
            \node[font=\footnotesize,text width=30pt,align=right,anchor=north east] at (340.1435684,789137.56) {single-ciphertext};
            \node[font=\footnotesize,text width=30pt,align=left,anchor=north west] at (385.3431673,789137.56) {multi-ciphertext};
            \draw[->] (315.1729698,789137.56) -- (207.9366135,789137.56);
            \draw[->] (415.8732269,789137.56) -- (630.3459396,789137.56);
        
            \draw[very thin, dashed] (axis cs:2896.30937574, 26.80213093) -- (axis cs:2896.30937574, 1288135.198) node[pos=0.14,left,rotate=90,anchor=north,font=\footnotesize] {CPU limit};

            \end{axis}
        \end{tikzpicture}
        \caption{Sorting}
        \label{fig:runtime:sort}
    \end{subfigure}
    \caption{Runtime of ranking, minimum, median, and sorting for increasing vector size. Related work's performance is reported as baseline: Phoenix~\cite{jovanovic2022private} and NEXUS~\cite{zhang2024secure} for minimum, and Hong et al.~\cite{hong2021efficient} for median and sorting. All solutions are assessed both in single-threaded (ST, dashed lines) and multi-threaded (MT, solid lines) settings. Both axes are in logarithmic scale.}
    \label{fig:runtime}
\end{figure*}
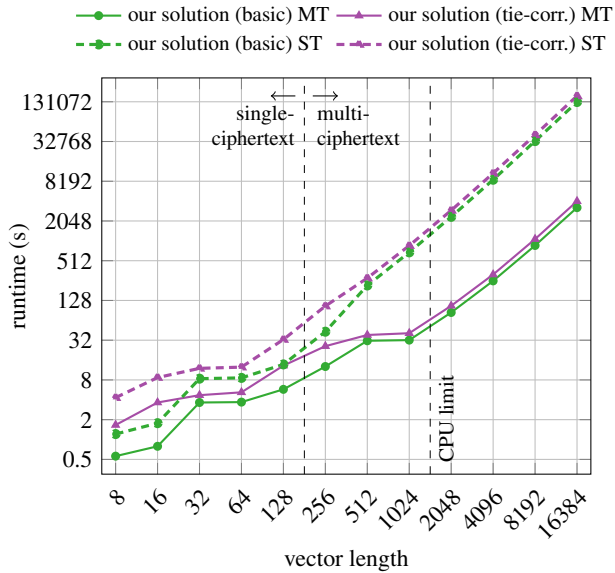
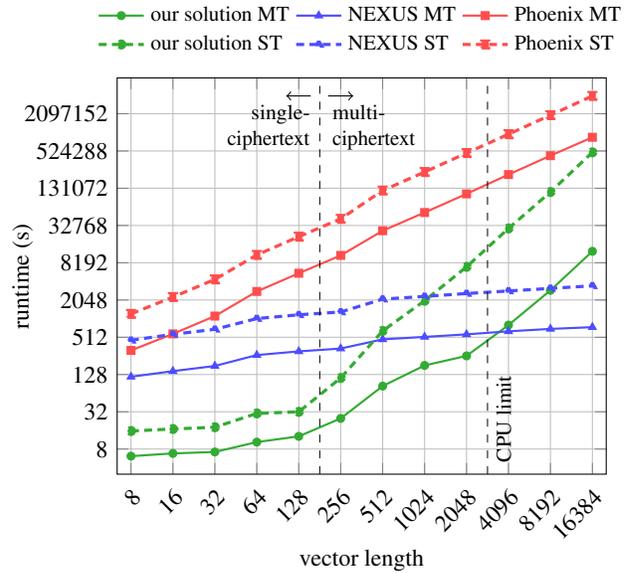
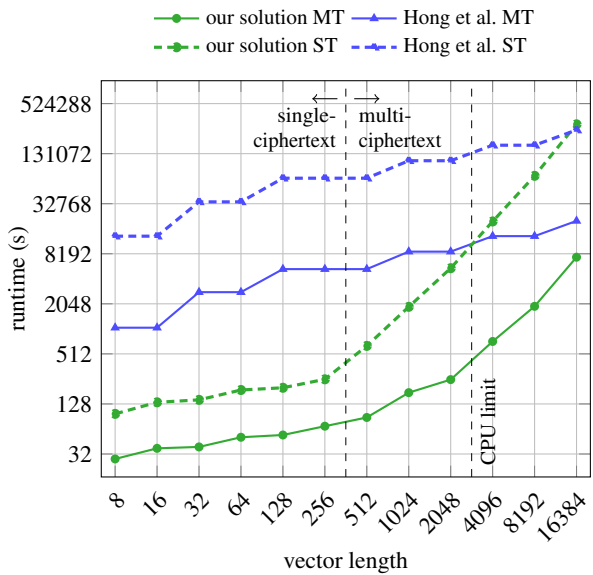
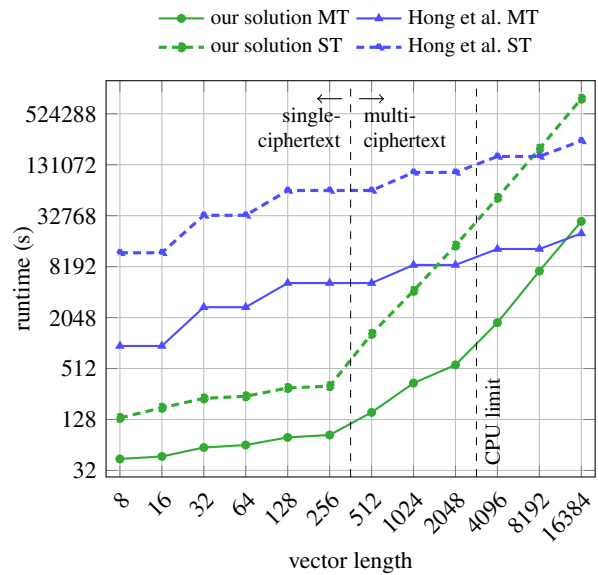

\begin{table*}[!b]
    \centering
    \caption{Memory consumption comparison of different solutions, both in single-thread (ST) and multi-thread (MT).}
    \label{tab:memory}
    \begin{subtable}[t]{0.45\textwidth}
        \centering
        \caption{Ranking}
        \small
        \begin{tabular}{rrrrr}
            \toprule
            & \multicolumn{2}{r}{\normalsize Our Solution (basic)} & \multicolumn{2}{r}{\normalsize Our Solution (tie corr.)} \\
            \normalsize $\vectorLength$ & \normalsize ST & \normalsize MT & \normalsize ST & \normalsize MT \\
            \midrule
            8 & 191 MB & 191 MB & 514 MB & 508 MB \\
            16 & 271 MB & 259 MB & 739 MB & 748 MB \\
            32 & 765 MB & 781 MB & 952 MB & 961 MB \\
            64 & 881 MB & 891 MB & 1.06 GB & 1.07 GB \\
            128 & 1.05 GB & 1.06 GB & 1.42 GB & 1.43 GB \\
            256 & 1.15 GB & 1.58 GB & 1.54 GB & 2.21 GB \\
            512 & 4.08 GB & 6.77 GB & 4.67 GB & 7.63 GB \\
            1024 & 4.43 GB & 13.3 GB & 5.13 GB & 15.4 GB \\
            2048 & 5.09 GB & 21.0 GB & 6.04 GB & 24.4 GB \\
            4096 & 6.55 GB & 22.4 GB & 8.01 GB & 26.6 GB \\
            8194 & 9.95 GB & 25.1 GB & 12.6 GB & 30.9 GB \\
            16384 & 18.8 GB & 30.8 GB & 25.0 GB & 40.2 GB \\
            \bottomrule
        \end{tabular}
        \vspace{10pt}
    \end{subtable}
    \hspace{20pt}
    \begin{subtable}[t]{0.45\textwidth}
        \centering
        \caption{Minimum}
        \small
        \begin{tabular}{rrrrr}
            \toprule
            & \multicolumn{2}{r}{\normalsize Our Solution} & \multicolumn{2}{r}{\normalsize NEXUS~\cite{zhang2024secure}} \\
            \normalsize $\vectorLength$ & \normalsize ST & \normalsize MT & \normalsize ST & \normalsize MT \\
            \midrule
            8 & 1.34 GB & 1.32 GB & 9.65 GB & 9.68 GB \\
            16 & 1.61 GB & 1.64 GB & 11.0 GB & 11.2 GB \\
            32 & 1.91 GB & 1.93 GB & 13.7 GB & 14.6 GB \\
            64 & 2.60 GB & 2.62 GB & 15.0 GB & 15.6 GB \\
            128 & 2.93 GB & 2.95 GB & 15.8 GB & 15.9 GB \\
            256 & 3.47 GB & 4.94 GB & 18.6 GB & 19.8 GB \\
            512 & 18.9 GB & 23.2 GB & 34.1 GB & 36.3 GB \\
            1024 & 20.0 GB & 39.0 GB & 35.3 GB & 37.7 GB \\
            2048 & 22.6 GB & 99.6 GB & 37.7 GB & 39.4 GB \\
            4096 & 30.4 GB & 113 GB & 60.2 GB & 64.3 GB \\
            8194 & 34.4 GB & 126 GB & 60.9 GB & 65.0 GB \\
            16384 & 46.4 GB & 165 GB & 62.2 GB & 66.3 GB \\
            \bottomrule
        \end{tabular}
    \end{subtable}
    \begin{subtable}[t]{0.45\textwidth}
        \centering
        \caption{Median}
        \small
        \begin{tabular}{rrrrr}
            \toprule
            & \multicolumn{2}{r}{\normalsize Our Solution} & \multicolumn{2}{r}{\normalsize Hong et al.~\cite{hong2021efficient}} \\
            \normalsize $\vectorLength$ & \normalsize ST & \normalsize MT & \normalsize ST & \normalsize MT \\
            \midrule
            8 & 5.73 GB & 5.86 GB & 12.9 GB & 37.9 GB \\
            16 & 7.89 GB & 8.02 GB & 12.9 GB & 37.9 GB \\
            32 & 9.42 GB & 9.52 GB & 12.9 GB & 37.9 GB \\
            64 & 12.2 GB & 12.3 GB & 12.9 GB & 37.9 GB \\
            128 & 13.9 GB & 14.1 GB & 12.9 GB & 37.9 GB \\
            256 & 16.9 GB & 17.2 GB & 12.9 GB & 37.9 GB \\
            512 & 18.7 GB & 23.0 GB & 12.9 GB & 37.9 GB \\
            1024 & 21.9 GB & 35.2 GB & 12.9 GB & 37.9 GB \\
            2048 & 25.6 GB & 65.1 GB & 12.9 GB & 37.9 GB \\
            4096 & 29.8 GB & 111 GB & 12.9 GB & 37.9 GB \\
            8194 & 34.6 GB & 129 GB & 12.9 GB & 37.9 GB \\
            16384 & 40.0 GB & 192 GB & 12.9 GB & 37.9 GB \\
            \bottomrule
        \end{tabular}
    \end{subtable}
    \hspace{20pt}
    \begin{subtable}[t]{0.45\textwidth}
        \centering
        \caption{Sorting}
        \small
        \begin{tabular}{rrrrr}
            \toprule
            & \multicolumn{2}{r}{\normalsize Our Solution} & \multicolumn{2}{r}{\normalsize Hong et al.~\cite{hong2021efficient}} \\
            \normalsize $\vectorLength$ & \normalsize ST & \normalsize MT & \normalsize ST & \normalsize MT \\
            \midrule
            8 & 7.22 GB & 7.34 GB & 10.9 GB & 35.9 GB \\
            16 & 8.95 GB & 9.03 GB & 10.9 GB & 35.9 GB \\
            32 & 11.6 GB & 11.7 GB & 10.9 GB & 35.9 GB \\
            64 & 13.5 GB & 13.7 GB & 10.9 GB & 35.9 GB \\
            128 & 16.8 GB & 17.0 GB & 10.9 GB & 35.9 GB \\
            256 & 18.8 GB & 19.0 GB & 10.9 GB & 35.9 GB \\
            512 & 21.7 GB & 28.0 GB & 10.9 GB & 35.9 GB \\
            1024 & 23.2 GB & 42.2 GB & 10.9 GB & 35.9 GB \\
            2048 & 26.9 GB & 97.0 GB & 10.9 GB & 35.9 GB \\
            4096 & 31.2 GB & 116 GB & 10.9 GB & 35.9 GB \\
            8194 & 36.0 GB & 155 GB & 10.9 GB & 35.9 GB \\
            16384 & 41.3 GB & 230 GB & 10.9 GB & 35.9 GB \\
            \bottomrule
        \end{tabular}
    \end{subtable}
\end{table*}

We proceed on the same line to adapt sorting to the multi-ciphertext setting.
First, a multi-ciphertext ranking is computed.
As we have to extract $\vectorLength$ order statistics, and each one could be in any of the ciphertext blocks, the ranking blocks are replicated both row-wise and block-wise.
Then each row of each block is shifted by a constant going from $1$ to $\vectorLength$, as in \Cref{alg:sorting}, although this time it spans over multiple instances of the same ranking block.
We conclude by applying the indicator function to each instance of each ranking block and summing up the results, both row- and block-wise.
A detailed description is presented in \Cref{alg:multiCtxt_sorting}.


\section{Experimental Evaluation}

We evaluate the performance of our designs for different vector sizes and compare them with existing work.

\subsection{Experimental Setup}

We use the CKKS implementation provided by the OpenFHE library~\cite{openFHE},\footnote{\url{https://github.com/openfheorg/openfhe-development}} with a scaling factor (decimal precision) ranging from 30 to 59 bits.
The ring dimension goes up to $2^{16}$ for ranking, and $2^{17}$ for order statistics and sorting, to accommodate the higher multiplicative depth.
The parameters are chosen in accordance with the Homomorphic Encryption Standard to assure 128-bit security~\cite{albrecht2015concrete, HomomorphicEncryptionSecurityStandard}.
Our code is available at \url{https://github.com/FedericoMazzone/openfhe-statistics}.

We present the runtime of ranking, computing the minimum, median, and sorting elements that are generated uniformly at random in a bounded interval.
For ranking and minimum, we use Chebyshev approximation of the comparison function up to degree $2^{11}$ for $\vectorLength \le 256$, while we employ the $f,g$ approximation by Cheon et al.~\cite{cheon2020efficient} for $\vectorLength > 256$, as we start benefiting from the lower runtime.
The composition degrees used are $d_f = 2$ and $d_g = 3$, which are the same we adopt in the median and sorting experiments for any $\vectorLength$.

As the depth of our circuit is upper-bounded by 65, bootstrapping is not needed, and the scheme is used as a leveled homomorphic encryption.
All the experiments are performed on a Linux machine with Intel Xeon Platinum 8358 running at 2.60 GHz, with 32 cores (64 threads), and 512 GB RAM.

\subsection{Empirical Results}

In \Cref{fig:runtime}, we report the runtime performance of our solution for the different functionalities described in \Cref{sec:main-design}, both in single-threaded (ST) and multi-threaded (MT) settings, with a maximum of 64 concurrent threads.

Our approach demonstrates particularly fast performance for small-sized input vectors that can be processed within a single ciphertext.
For example, computing the minimum of a vector takes only 15.65s for $\vectorLength = 8$, and up to 31.95s for $\vectorLength = 128$, in ST.
However, as soon as the input vector needs to be split into multiple chunks --- of 128 elements for ranking and minimum, and of 256 elements for median and sorting --- our solution exhibits a steep increase in runtime.
The runtime for the minimum computation jumps to 112.61s for 2 chunks (256 elements), 648.55s for 4 chunks (512 elements), 1994.30s for 8 chunks (1024 elements), and so on.

This slow-down due to switching from single- to multi-ciphertext mode can be observed in all four functionalities.
The primary cause is the increased number of comparisons required in multi-ciphertext mode.
This effect is particularly noticeable in the ST setting, where comparisons are evaluated sequentially, making the quadratic growth of our solution's cost evident in the runtime.
In the MT setting, the slow-down becomes more pronounced once we exceed the available CPU threads (CPU limit in \Cref{fig:runtime}) and can no longer parallelize the comparison evaluations.
Specifically, this occurs when the number of blocks increases from 8 to 16, causing the number of comparisons to grow from 36 to 136, which exceeds the 64-thread capacity of our machine.
Beyond this point, comparisons begin to execute sequentially, which has a significant impact on the runtime of our solution.

In \Cref{fig:runtime:rank} we can see the computational overhead introduced by the tie-correction offset in the ranking algorithm.
In the MT setting, ranking takes 0.56s for $\vectorLength = 8$ and 12.77s for $\vectorLength = 256$, while the tie-correction increases these runtimes to 1.66s and 25.97s, respectively.
The overhead varies within this range, peaking at 13.2s, while it stabilizes at around 25–27\% for $\vectorLength > 256$.
Note that this overhead is not only due to the extra operations required to compute the correction offset, but also to the increased multiplicative depth, which makes all operations computationally more expensive.

In \Cref{fig:runtime:median} and \Cref{fig:runtime:sort} we report the runtime for the median and sorting algorithms, respectively.
We will discuss them in more detail in \Cref{sec:comparison-with-previous-work}.
For the moment, we only highlight that the median serves as a representative test for order statistic extraction with tie-correction enabled.
The cost of computing any other order statistic is identical, as it only changes the interval of the indicator function.

Finally, \Cref{tab:memory} shows that memory consumption grows steadily with the input size, and the switch from single- to multi-ciphertext has a relatively minor impact compared to the jumps caused by an increase in the ring dimension.
For example, in the case of ranking (basic) in a single-thread setting, transitioning from single-ciphertext ($N=128$) to multi-ciphertext ($N=256$) results in a modest memory increase from 1.05GB to 1.15GB (+10\%).
In contrast, when the ring dimension increases, the memory consumption grows significantly: from $N=16$ to $N=32$ (ring dimension: $2^{14} \to 2^{15}$), the memory rises from 271MB to 765MB (+182\%), and from $N=256$ to $N=512$ (ring dimension: $2^{15} \to 2^{16}$), it increases from 1.15GB to 4.08GB (+255\%).
This is expected since each ciphertext requires at least twice as many bits to represent after a ring dimension increase.

\subsection{Comparison with Previous Work}
\label{sec:comparison-with-previous-work}

We compare our approach to state-of-the-art solutions for computing the minimum/maximum, median, and sorting. To ensure a fair comparison, for each experiment we use the same comparison approximation method and degree across all evaluated solutions.

\paragraph{Minimum and Maximum}
For computing the minimum and maximum, we assess our approach against two existing solutions: Phoenix\cite{jovanovic2022private} and NEXUS\cite{zhang2024secure}.
Like our solution, both Phoenix and NEXUS operate on elements encrypted within a single ciphertext.
\Cref{tab:cmp_with_prev_work:order_statistics} provides a detailed comparison, focusing on the number of evaluations of the comparison function, homomorphic rotations, and the number of slots required in the ciphertext.
While the logarithmic and linear scaling of NEXUS and Phoenix, respectively, enable their solutions to handle larger input vectors more efficiently, our approach performs better for small input sizes.
As shown in \Cref{fig:runtime:min}, our solution is faster for $\vectorLength \leq 1024$ in single-threaded (ST) settings and $\vectorLength \leq 2048$ in multi-threaded (MT) settings, given our hardware configuration with 64 threads.
For larger input vectors, NEXUS becomes the preferred choice due to its superior scalability.
By design, Phoenix consistently lags behind NEXUS in runtime performance.

When the input vector is relatively short, our approach provides a significant speed-up over existing solutions.
For example, in the use case of NEXUS~\cite{zhang2024secure}, where the argmax is applied for secure transformer inference, particularly for computing the output layer in BERT-based and GPT-2 models with $\vectorLength=128$ nodes, we observe the following:
\begin{itemize}
    \item Phoenix~\cite{jovanovic2022private} requires $128$ comparisons, $128$ rotations, resulting in a total runtime of 92.31 minutes;
    \item NEXUS~\cite{zhang2024secure} requires $7$ comparisons, $7$ rotations, with a total runtime of 5.05 minutes;
    \item our approach requires $2$ comparisons, $28$ rotations, resulting in a total runtime of 12.83 seconds.
\end{itemize}
However, it is important to note that NEXUS and Phoenix have different space requirements compared to our approach, as summarized in \Cref{tab:cmp_with_prev_work:order_statistics}.
This enables them to utilize extra space for batching computations.
For instance, if a ciphertext can encode 16,384 elements, our solution processes only 1 argmax in this example, while NEXUS and Phoenix can process 64 and 128 vectors simultaneously, respectively.

Memory usage follows a similar trend, as shown in \Cref{tab:memory}.
Our approach consumes less memory than NEXUS for small inputs, but the memory consumption increases significantly with larger inputs.
The memory consumption of Phoenix is not explicitly reported as it is equivalent to that of NEXUS.

\begin{table}[b]
    \caption{Summary of different solutions for computing the minimum and maximum functionalities.}
    \label{tab:cmp_with_prev_work:order_statistics}
    \centering
    \begin{tabular}{llll}
        \toprule
        & \thead{Comparisons} & \thead{Rotations} & \thead{Slots} \\
        \midrule
        Phoenix~\cite{jovanovic2022private} & $O(\vectorLength)$ & $O(\vectorLength)$ & $\vectorLength$ \\
        NEXUS~\cite{zhang2024secure} & $O(\log{\vectorLength})$ & $O(\log{\vectorLength})$ & $2 \vectorLength$ \\
        Our work & $O(\numBlocks^2)$ & $O(\log{\vectorLength})$ & $\vectorLength^2$ \\
        \bottomrule
    \end{tabular}
\end{table}

\paragraph{Median and Sorting}
For sorting vectors, we compare our approach with the state-of-the-art for CKKS, that is the k-way sorting network approach by Hong et al.~\cite{hong2021efficient}.
While their method also leverages the SIMD capabilities of the encryption scheme, it results in a comparison depth of $k \log_k^2{\vectorLength}$ when employing a k-way network.
Their approach performs a total of $O(\vectorLength \log_k^2{\vectorLength})$ comparisons, making it more scalable than our solution for larger input vectors.
This scalability advantage is evident in \Cref{fig:runtime:sort}, where we report their runtime for $k = 5$, the optimal parameter choice for their solution.
However, our approach outperforms Hong et al.~\cite{hong2021efficient} for input sizes up to $\vectorLength = 4096$ in ST settings, and up to $\vectorLength = 8192$ in MT settings.

We extended Hong et al.'s scheme to compute the median. Specifically, we sort the input vector using their algorithm, extract the values at indices $\vectorLength/2$ and $\vectorLength/2 + 1$, and then compare these median values with the original vector to determine the median indices.
As shown in \Cref{fig:runtime:median}, the overhead introduced by this extension is minimal.
Nonetheless, combined with the slightly lower computational cost of our median algorithm compared to full sorting, it shifts the input size thresholds where our solution is beneficial over theirs to $\vectorLength = 8192$ in ST and $\vectorLength = 16384$ in MT.
In terms of memory usage, we note that their solution maintains constant memory consumption, whereas our approach scales with the input size.

\subsection{Applications and Limitations}

The main limitation of our approach resides in its quadratic space complexity, which forces us to split the input vector in many smaller chunks, quickly increasing the number of necessary comparison evaluations.
This is relevant especially for large inputs or in environments with limited parallelization capabilities.
As a result, the scalability of our solution is hindered in such scenarios.

However, our solution is well-suited for applications that do not involve processing large vectors.
For example, it may be effective in scenarios involving outsourced data analysis of small datasets, which are often encountered in healthcare studies where hospitals analyze data from hundreds of patients.
Additionally, it could perform well on datasets with categorical attributes. Those could be represented in a one-hot encoding, making categorical operations (e.g., mode) straightforward to implement within our approach. In such cases, as categorical attributes typically have a limited number of possible values, we would be dealing with short vectors.

In the context of privacy-preserving machine learning (PPML), our approach can be applied in secure inference tasks. For instance, it can be useful for computing the max-pooling layers of a convolutional neural network (CNN), where the maximum is calculated over a kernel-sized vector, typically 3x3, 5x5, or 7x7. It is also applicable in extracting the argmax from the output layer of most neural networks for classification tasks. In such cases, the vector size corresponds to the number of classes in the classification problem, which is typically in the range of 2 to 100, depending on the application.

Beyond inference, in the context of PPML training, we foresee that our solution could be employed to securely train simple unsupervised models in federated settings. A notable example is k-means clustering, where the main computation involves finding the argmin of distances over k clusters. Here, k often takes values such as 2, 5, or 8, making our approach particularly suitable for such tasks.


\section{Related Work}


A vast body of literature has focused on either sorting elements or computing their maximum value under encryption.
Sorting under FHE has been studied starting from 2010 under the Smart-Vercauteren (SV) scheme~\cite{smart2010fully} using bitwise encoding and comparison based swaps to implement algorithms like Bubble Sort, Insertion Sort~\cite{chatterjee2013accelerating}, and Quick Sort~\cite{chatterjee2017sorting}, all of which requiring $O(\vectorLength^2)$ comparisons.
Subsequently, Bitonic Sort and Odd-Even Merge Sort were also implemented, reducing the cost to $O(\vectorLength \log^2{\vectorLength})$ comparisons, of which $\vectorLength$ can be potentially run in parallel, making the comparison depth $\log^2{\vectorLength}$ ~\cite{emmadi2015updates}.
In 2021, some works started designing sorting for floating-point values under CKKS.
Hong et al.~\cite{hong2021efficient} use k-way sorting networks to achieve a $k \log_k^2{\vectorLength}$ comparison depth.
While Lu et al. (PEGASUS)~\cite{lu2021pegasus} also implement Bitonic Sort but performing the comparisons using the efficient look-up tables of FHEW~\cite{ducas2015fhew} after a scheme switching from CKKS.

An entire line of work has focused specifically on improving on the evaluation of the comparison function itself.
Chialva et al.~\cite{chialva2018conditionals} use the identity $\tanh(kx) = (e^{kx} - e^{-kx}) / (e^{kx} + e^{-kx})$ to approximate the sign function for large $k > 0$,
while Boura et al.~\cite{boura2020chimera} employ an approximation based on Fourier series.
The work by Cheon et al.~\cite{cheon2019numerical} is the first one to study the max function under CKKS, and it is based on an iterative computation of $u^k / (u^k + v^k)$ for large $k > 0$.
The same author proposes a new solution in~\cite{cheon2020efficient}, where a composition of 2 polynomials $f,g$ is used to approximate the sign function, proving an optimal asymptotic complexity.
This study was then picked up by Lee et al.~\cite{lee2021minimax}, who generalized the technique to composition of $k$ polynomials, and found the optimal set of polynomials for any given multiplicative depth.

In Phoenix~\cite{jovanovic2022private}, the authors face the problem of computing the argmax in the output layer of a neural network to perform privacy-preserving inference.
There, the elements are stored within a single CKKS ciphertext and they propose a method based on rotations to compute the argmax in $\vectorLength$ comparisons and $\vectorLength$ rotations.
In NEXUS~\cite{zhang2024secure}, the authors apply the same strategy recursively, exploiting SIMD slot folding, which results in comparing the elements in a binary tree fashion, thus reducing the cost to $\log{\vectorLength} + 1$ rotations and $\log{\vectorLength} + 1$ comparisons.
They use it for secure transformer inference, in particular for computing the argmax output layer in BERT-based and GPT-2 models.

It is also worth mentioning the work by Lu et al.~\cite{lu2016using}, where the authors propose FHE algorithms that compute a variety of descriptive statistics, including percentile.
However, their method is limited to ordinal attributes and requires plaintext encoding dependent on value order, whereas our approach addresses numerical attributes, operating with encrypted vectors without specific plaintext encoding, and thus it can be easily integrated in larger (numerical) circuits.
Our paper contributes to these lines of work by introducing a novel approach for implementing comparison-based functionalities that achieve a constant comparison depth of $2$.
This represents an important reduction with respect to existing solutions, which have higher comparison depths, as also summarized in \Cref{tab:related_work}.


\section{Conclusion}

In this paper, we have presented a novel approach for computing ranking, order statistics, and sorting of a vector under CKKS.
Our method relies on homomorphic matrix encoding and on the SIMD capabilities of the cryptosystem to compare all elements with each other at once, reducing the comparison depth of these algorithms to 2.
This makes our solution highly parallelizable, opening potential future work in the direction of hardware acceleration.
We showed that our approach is beneficial over existing solutions when the input vector is within the order of thousand of elements, achieving remarkable speed-ups, and particularly shining in multi-threaded settings.
We consider the algorithms we designed practical for a wide range of privacy-preserving scenarios, especially for data outsourcing and secure machine learning, or for serving as fundamental building blocks for larger protocols.


\section*{Acknowledgment}
This project has received funding from the European Union’s Horizon 2020 research and innovation programme under Grant Agreement No 965315. The results reflect only the authors' view and the European Commission is not responsible for any use that may be made of the information this paper contains.
This work was also supported by the Netherlands Organization for Scientific Research under NWO:SHARE project [CS.011].





\section*{Ethics Considerations}

Our work focuses solely on computational methods under encryption, and no real-world data has been used to test our approach.
Consequently, we see no privacy concerns, risks of data misuse, or potential harm to individuals or communities arising from our work.
The algorithms and protocols we developed are purely theoretical in nature, designed to enhance computational efficiency and security in encrypted environments.
They do not interact with human subjects or physical systems in any way that could cause harm or raise ethical concerns.
Our methods can be actually put in place to defend sensitive data in specific applications.


\section*{Open Science}


In accordance with the principles of Open Science, we have made the complete codebase associated with this paper publicly available under the BSD 2-Clause license.
The codebase represents the sole artifact accompanying this work and includes the implementation of all functionalities described in the paper.
Specifically, the code consists of a C++ library built on top of the OpenFHE library.
The permanent link to the artifact is: \url{https://doi.org/10.5281/zenodo.14673904}.
The primary components (source files) correspond to the key functionalities discussed in Section~\ref{sec:main-design}:
\begin{enumerate}
    \item ranking,
    \item minimum,
    \item median,
    \item sorting.
\end{enumerate}

Additionally, for each functionality, we provide a benchmarking script that measures runtime performance of our solution on randomly generated vectors under various settings.
These scripts were used to produce the runtime results reported in Figure~\ref{fig:runtime}, which represents the main experimental assessment of our work.
Each script takes as input the vector length and the option to run in either single-threaded or multi-threaded mode.
The memory consumption data presented in Table~\ref{tab:memory} was collected using the Linux top command during these benchmark executions.


\bibliographystyle{plain}
\bibliography{bibliography}



\appendix

\section{Recursive Matrix Operations}
\label{app:rec_matrix_ops}

We provide the pseudocode for $\sumR$, $\sumC$, $\replR$, $\replC$ for a square matrix with $\vectorLength$ number of rows/columns. The matrix is assumed to be padded in such a way that $\vectorLength$ is a power of 2.

\begin{algorithm}[!htb]
\caption{\sumR}
\label{alg:sumR}
\begin{algorithmic}[1]
\Require $X$ encryption of a square matrix of size $\vectorLength$.
\Ensure $X$ encryption of a row vector.
\For{$i = 0, \dots, \log{\vectorLength} - 1$}
    \State $X \gets X + (X \ll \vectorLength \cdot 2^i)$
\EndFor
\State $X \gets\maskR(X, 0)$
\State \Return $X$
\end{algorithmic}
\end{algorithm}

\begin{algorithm}[!htb]
\caption{\sumC}
\label{alg:sumC}
\begin{algorithmic}[1]
\Require $X$ encryption of a square matrix of size $\vectorLength$.
\Ensure $X$ encryption of a column vector.
\For{$i = 0, \dots, \log{\vectorLength} - 1$}
    \State $X \gets X + (X \ll 2^i)$
\EndFor
\State $X \gets\maskC(X, 0)$
\State \Return $X$
\end{algorithmic}
\end{algorithm}

\begin{algorithm}[!htb]
\caption{\replR}
\label{alg:replR}
\begin{algorithmic}[1]
\Require $X$ encryption of a row vector of size $\vectorLength$.
\Ensure $X$ encryption of a square matrix.
\For{$i = 0, \dots, \log{\vectorLength} - 1$}
    \State $X \gets X + (X \gg \vectorLength \cdot 2^i)$
\EndFor
\State \Return $X$
\end{algorithmic}
\end{algorithm}

\begin{algorithm}[!htb]
\caption{\replC}
\label{alg:replC}
\begin{algorithmic}[1]
\Require $X$ encryption of a column vector of size $\vectorLength$.
\Ensure $X$ encryption of a square matrix.
\For{$i = 0, \dots, \log{\vectorLength} - 1$}
    \State $X \gets X + (X \gg 2^i)$
\EndFor
\State \Return $X$
\end{algorithmic}
\end{algorithm}


\section{Effect of Chebyshev Approximation Degree on Performance}
\label{app:effect_approx_degree}

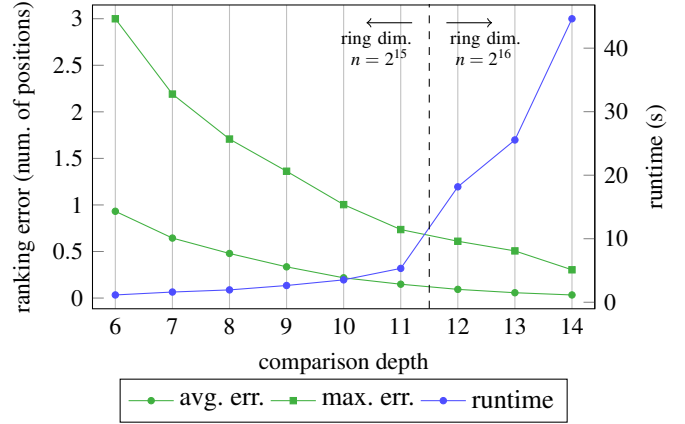
\begin{figure}[!htb]
    \centering
    \begin{tikzpicture}
    \begin{axis}[
        width=235pt,
        height=160pt,
        xlabel=comparison depth,
        ylabel=ranking error (num. of positions),
        xtick pos=bottom,
        ytick pos=left,
        xmajorgrids,
        xtick=data,
        ytick={0,0.5,...,3},
        xtick style={draw=none},
        enlarge x limits=0.05,
        enlarge y limits=0.05,
        label style={font=\small},
        tick label style={font=\small}
        ]
    \addplot[color=supergreen!75,mark=*,mark size=1.2pt]
        plot coordinates {
            ( 6, 0.931911)
            ( 7, 0.644247)
            ( 8, 0.47884)
            ( 9, 0.3353)
            (10, 0.217489)
            (11, 0.148662)
            (12, 0.0933653)
            (13, 0.0571904)
            (14, 0.0333236)
        };\label{plot_avg_err}

    \addplot[color=supergreen!75,mark=square*,mark size=1.2pt]
        plot coordinates {
            ( 6, 2.9993)
            ( 7, 2.19057)
            ( 8, 1.70762)
            ( 9, 1.36205)
            (10, 1.00342)
            (11, 0.735678)
            (12, 0.610084)
            (13, 0.506405)
            (14, 0.303459)
        };\label{plot_max_err}
    \end{axis}
    
    \begin{axis}[
        width=235pt,
        height=160pt,
        axis x line=none,
        axis y line*=right,
        ylabel=runtime (s),
        legend style={at={(0.50,-0.23)},anchor=north},
        legend columns=3,
        legend cell align=left,
        ytick pos=right,
        xmajorgrids,
        ytick={0,10,20,30,40},
        xtick style={draw=none},
        enlarge x limits=0.05,
        enlarge y limits=0.05,
        label style={font=\small},
        tick label style={font=\small}
        ]
    \addlegendimage{/pgfplots/refstyle=plot_avg_err}\addlegendentry{avg. err.}
    \addlegendimage{/pgfplots/refstyle=plot_max_err}\addlegendentry{max. err.}
    \addplot[color=blue!70,mark=*,mark size=1.2pt]
        plot coordinates {
            ( 6, 1.14823)
            ( 7, 1.60821)
            ( 8, 1.94349)
            ( 9, 2.63523)
            (10, 3.52464)
            (11, 5.33022)
            (12, 18.1701)
            (13, 25.5488)
            (14, 44.6263)
        };
    \addlegendentry{runtime}

    \draw[dashed] (axis cs:11.5, -1) -- (axis cs:11.5, 47);
    \node[font=\scriptsize,text width=30pt,align=right,anchor=east] at (11.3,40) {ring dim. $n = 2^{15}$};
    \node[font=\scriptsize,text width=30pt,align=center,anchor=west] at (11.6,40) {ring dim. $n = 2^{16}$};
    \draw[->] (11.2,44) -- (10.4,44);
    \draw[->] (11.8,44) -- (12.6,44);
    
    \end{axis}
    
    \end{tikzpicture}
    \caption{Ranking a vector of 128 elements for different approximation degrees of the comparison function (as multiplicative depth). The ranking error and runtime are reported.}
    \label{fig:rank_degree_vs_error}
\end{figure}

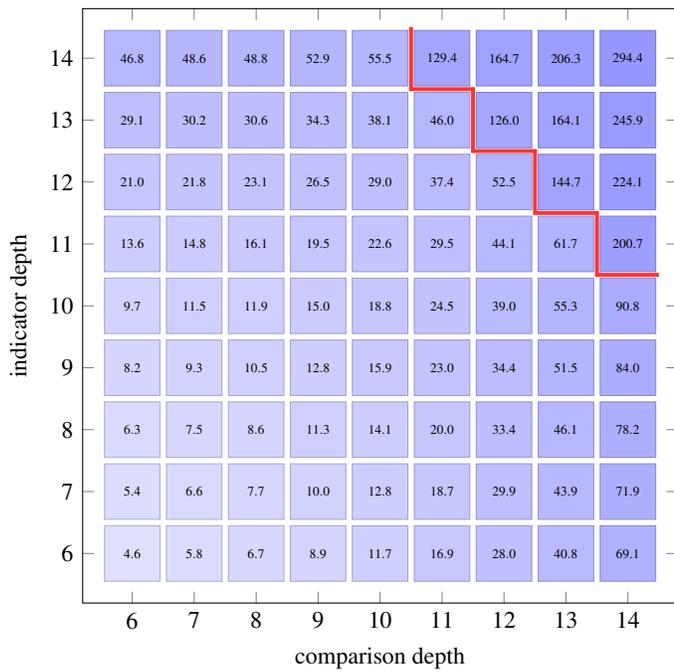
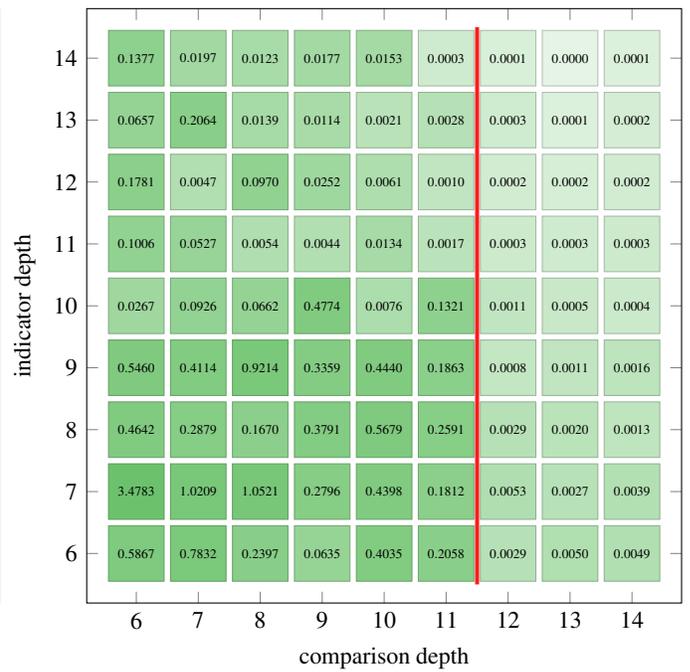
\begin{figure*}[!htb]
    \begin{subfigure}{.49\textwidth}
        \centering
        \begin{tikzpicture}
        \pgfplotstableread{
        X   Y   Runtime
        6	6	4.61997
        6	7	5.43719
        6	8	6.29068
        6	9	8.1623
        6	10	9.70991
        6	11	13.6112
        6	12	20.9793
        6	13	29.1406
        6	14	46.7658
        7	6	5.75086
        7	7	6.5839
        7	8	7.52977
        7	9	9.3026
        7	10	11.4671
        7	11	14.765
        7	12	21.8037
        7	13	30.16
        7	14	48.6383
        8	6	6.699
        8	7	7.70612
        8	8	8.63467
        8	9	10.5452
        8	10	11.9099
        8	11	16.0653
        8	12	23.055
        8	13	30.6453
        8	14	48.8381
        9	6	8.86585
        9	7	9.97534
        9	8	11.3049
        9	9	12.8494
        9	10	15.0045
        9	11	19.484
        9	12	26.463
        9	13	34.2968
        9	14	52.9082
        10	6	11.7213
        10	7	12.8257
        10	8	14.1499
        10	9	15.9444
        10	10	18.7892
        10	11	22.6308
        10	12	29.0468
        10	13	38.0629
        10	14	55.4536
        11	6	16.8914
        11	7	18.7432
        11	8	19.9853
        11	9	23.0017
        11	10	24.4806
        11	11	29.5267
        11	12	37.3507
        11	13	45.9694
        11	14	129.43
        12	6	27.9983
        12	7	29.8849
        12	8	33.3529
        12	9	34.3843
        12	10	39.0009
        12	11	44.1445
        12	12	52.5084
        12	13	125.985
        12	14	164.73
        13	6	40.7799
        13	7	43.8708
        13	8	46.0673
        13	9	51.4726
        13	10	55.2874
        13	11	61.7336
        13	12	144.712
        13	13	164.11
        13	14	206.298
        14	6	69.076
        14	7	71.8542
        14	8	78.1852
        14	9	84.0409
        14	10	90.801
        14	11	200.747
        14	12	224.064
        14	13	245.881
        14	14	294.36
        }\data
        
        \pgfplotsset{
            colormap={custom}{
                rgb=(0.9,0.9,1)
                rgb=(0.8,0.8,1)
                rgb=(0.7,0.7,1)
                rgb=(0.6,0.6,1)
                rgb=(0.5,0.5,1)
                rgb=(0.4,0.4,1)
                rgb=(0.3,0.3,1)
                rgb=(0.2,0.2,1)
                rgb=(0.1,0.1,1)
                rgb=(0,0,1)
            },
        }
        
        \begin{axis}[
            width=270pt,
            height=270pt,
            xlabel=comparison depth,
            ylabel=indicator depth,
            xtick={6,7,8,9,10,11,12,13,14},
            ytick={6,7,8,9,10,11,12,13,14},
            point meta min=0.5,
            point meta max=6.25,
            label style={font=\small},
            tick label style={font=\small}
        ]
        
        \addplot[
            scatter,
            only marks,
            mark=square*,
            mark size=10.5pt,
            point meta={log10(\thisrow{Runtime})}, 
            visualization depends on=\thisrow{Runtime} \as \myvalue,
            nodes near coords*={\pgfmathprintnumber[fixed zerofill,precision=1]{\myvalue}},
            every node near coord/.append style={font=\tiny, anchor=center}
        ] table {\data};

        \draw[line width=1.5pt,red!80]
            (14.5,10.5) --
            (13.5,10.5) --
            (13.5,11.5) --
            (12.5,11.5) --
            (12.5,12.5) --
            (11.5,12.5) --
            (11.5,13.5) --
            (10.5,13.5) --
            (10.5,14.5);
        
        \end{axis}
        \end{tikzpicture}
        \caption{runtime in seconds}
        \label{fig:min_degree_vs_runtime_and_error:runtime}
    \end{subfigure}\hspace{8pt}
    \begin{subfigure}{.49\textwidth}
        \centering
        \begin{tikzpicture}
        \pgfplotstableread{
        X   Y   Error
        6	6	0.586741
        6	7	3.47832
        6	8	0.464231
        6	9	0.546033
        6	10	0.0266904
        6	11	0.100584
        6	12	0.178081
        6	13	0.0656801
        6	14	0.13766
        7	6	0.783171
        7	7	1.02091
        7	8	0.287859
        7	9	0.41143
        7	10	0.0925754
        7	11	0.052716
        7	12	0.00469829
        7	13	0.206426
        7	14	0.0196837
        8	6	0.239684
        8	7	1.05212
        8	8	0.166985
        8	9	0.921431
        8	10	0.0662125
        8	11	0.00537028
        8	12	0.0969621
        8	13	0.0138736
        8	14	0.0123461
        9	6	0.0634505
        9	7	0.279553
        9	8	0.379066
        9	9	0.335858
        9	10	0.477377
        9	11	0.00436281
        9	12	0.0251951
        9	13	0.0113654
        9	14	0.017687
        10	6	0.403522
        10	7	0.439828
        10	8	0.567901
        10	9	0.443957
        10	10	0.00757102
        10	11	0.0134407
        10	12	0.00606658
        10	13	0.00209837
        10	14	0.0153161
        11	6	0.205845
        11	7	0.181215
        11	8	0.259126
        11	9	0.186323
        11	10	0.132104
        11	11	0.00174293
        11	12	0.00102367
        11	13	0.00283556
        11	14	0.000282049
        12	6	0.00291922
        12	7	0.00527615
        12	8	0.00293025
        12	9	0.000843967
        12	10	0.00105363
        12	11	0.00033138
        12	12	0.000202474
        12	13	0.000252964
        12	14	0.000116116
        13	6	0.00500121
        13	7	0.00265771
        13	8	0.00195419
        13	9	0.00112145
        13	10	0.000510581
        13	11	0.000277119
        13	12	0.000153675
        13	13	7.59E-05
        13	14	3.45E-05
        14	6	0.00488624
        14	7	0.00394068
        14	8	0.00126197
        14	9	0.00155217
        14	10	0.000406302
        14	11	0.000337175
        14	12	0.000202035
        14	13	0.00016962
        14	14	7.86E-05
        }\data
        
        \pgfplotsset{
            colormap={custom}{
                color=(supergreen!10)
                color=(supergreen!90)
            },
        }
        
        \begin{axis}[
            width=270pt,
            height=270pt,
            xlabel=comparison depth,
            ylabel=indicator depth,
            xtick={6,7,8,9,10,11,12,13,14},
            ytick={6,7,8,9,10,11,12,13,14},
            point meta min=-4.5,
            point meta max=4,
            label style={font=\small},
            tick label style={font=\small}
        ]
        
        \addplot[
            scatter,
            only marks,
            mark=square*,
            mark size=10.5pt,
            point meta={log10(\thisrow{Error})}, 
            visualization depends on=\thisrow{Error} \as \myvalue,
            nodes near coords*={\pgfmathprintnumber[fixed, fixed zerofill, precision=4]{\myvalue}},
            every node near coord/.append style={font=\tiny, anchor=center}
        ] table {\data};

        \draw[line width=1.5pt,red!90]
            (11.5,5.5) --
            (11.5,14.5);
        
        \end{axis}
        \end{tikzpicture}
        \caption{Error}
        \label{fig:min_degree_vs_runtime_and_error:error}
    \end{subfigure}
    \caption{Computing the minimum of a vector of 32 elements for different approximation degrees of the comparison and indicator functions (as multiplicative depth). The runtime and L1 error are reported.}
    \label{fig:min_degree_vs_runtime_and_error}
\end{figure*}

In this work we employ a relatively basic implementation of comparison and indicator functions.
There is an entire body of literature that focuses specifically on this topic, which one may consider for real-life application, see for instance~\cite{cheon2020efficient,lee2021minimax}.
Nonetheless, we still provide some indication on how different approximation degrees influence the runtime of our algorithms, in exchange of having a more precise result.

We use ranking and minimum computation as case studies for this analysis.
\Cref{fig:rank_degree_vs_error} shows the impact of the approximation degree of the comparison function, represented in terms of its multiplicative depth, on runtime and approximation error in the ranking task.
Higher degrees yield lower error but incur longer runtimes.
The reported error is for a vector of 128 elements, indicating how many positions each element is ranked away, on average and in the worst case.
The steep increase after depth 11 is due to the fact that the ring dimension must increase to assure 128 bits of security, making the basic homomorphic operations more expensive.

Sweet spots in the trade-off can be noticed at depth 10 and 11, which corresponds to an approximation degree of $2^9$ and $2^{10}$, respectively.
At depth 10, the runtime is around 3.52 seconds, while the elements are ranked no more than 1 position away from their actual rank.
Notably, this error is proportional to the separation between elements; closely positioned elements are more susceptible to rank swapping.
One could willingly decide to use a lower approximation degree and exploit this effect to achieve a form of differential privacy.

Similar considerations can be applied to the minimum computation.
\Cref{fig:min_degree_vs_runtime_and_error} reports the error as the L1 distance between the computed and the expected minimum values.
We can see that a sensitive improvement in accuracy occurs when transitioning from comparison depth 11 to 12.
Moreover, we notice a sweet spot for the runtime when the sum of the approximation depths equals 24 (the upper-right diagonal), with the (12, 12) combination yielding the lowest error.

\end{document}